
%
\input amstex

%
%
%
%
%
%


\magnification=1200
\hsize=31pc
\vsize=55 truepc
\hfuzz=2pt
\vfuzz=4pt
\pretolerance=500
\tolerance=500
\parskip=0pt plus 1pt
\parindent=16pt
%

%
%
\font\fourteenrm=cmr10 scaled \magstep2
\font\fourteeni=cmmi10 scaled \magstep2
\font\fourteenbf=cmbx10 scaled \magstep2
\font\fourteenit=cmti10 scaled \magstep2
\font\fourteensy=cmsy10 scaled \magstep2

%
\font\large=cmbx10 scaled \magstep1

%
\font\sans=cmssbx10

%

%

%
\font\eightrm=cmr8
\font\eighti=cmmi8
\font\eightbf=cmbx8
\font\eightit=cmti8

\font\eightsy=cmsy8
\font\sixrm=cmr6
\font\sixi=cmmi6
\font\sixsy=cmsy6

%
\def\tenpoint{\def\rm{\fam0\tenrm}%
  \textfont0=\tenrm \scriptfont0=\sevenrm
                      \scriptscriptfont0=\fiverm
  \textfont1=\teni  \scriptfont1=\seveni
                      \scriptscriptfont1=\fivei
  \textfont2=\tensy \scriptfont2=\sevensy
                      \scriptscriptfont2=\fivesy
  \textfont3=\tenex   \scriptfont3=\tenex
                      \scriptscriptfont3=\tenex
  \textfont\itfam=\tenit  \def\it{\fam\itfam\tenit}%
  \textfont\slfam=\tensl  \def\sl{\fam\slfam\tensl}%
  \textfont\bffam=\tenbf  \scriptfont\bffam=\sevenbf
                            \scriptscriptfont\bffam=\fivebf
                            \def\bf{\fam\bffam\tenbf}%
  \normalbaselineskip=20 truept
  \setbox\strutbox=\hbox{\vrule height14pt depth6pt
width0pt}%
  \let\sc=\eightrm \normalbaselines\rm}
\def\eightpoint{\def\rm{\fam0\eightrm}%
  \textfont0=\eightrm \scriptfont0=\sixrm
                      \scriptscriptfont0=\fiverm
  \textfont1=\eighti  \scriptfont1=\sixi
                      \scriptscriptfont1=\fivei
  \textfont2=\eightsy \scriptfont2=\sixsy
                      \scriptscriptfont2=\fivesy
  \textfont3=\tenex   \scriptfont3=\tenex
                      \scriptscriptfont3=\tenex
  \textfont\itfam=\eightit  \def\it{\fam\itfam\eightit}%
  \textfont\bffam=\eightbf  \def\bf{\fam\bffam\eightbf}%
  \normalbaselineskip=16 truept
  \setbox\strutbox=\hbox{\vrule height11pt depth5pt width0pt}}
\def\fourteenpoint{\def\rm{\fam0\fourteenrm}%
  \textfont0=\fourteenrm \scriptfont0=\tenrm
                      \scriptscriptfont0=\eightrm
  \textfont1=\fourteeni  \scriptfont1=\teni
                      \scriptscriptfont1=\eighti
  \textfont2=\fourteensy \scriptfont2=\tensy
                      \scriptscriptfont2=\eightsy
  \textfont3=\tenex   \scriptfont3=\tenex
                      \scriptscriptfont3=\tenex
  \textfont\itfam=\fourteenit  \def\it{\fam\itfam\fourteenit}%
  \textfont\bffam=\fourteenbf  \scriptfont\bffam=\tenbf
                             \scriptscriptfont\bffam=\eightbf
                             \def\bf{\fam\bffam\fourteenbf}%
  \normalbaselineskip=24 truept
  \setbox\strutbox=\hbox{\vrule height17pt depth7pt width0pt}%
  \let\sc=\tenrm \normalbaselines\rm}
\def\today{\number\day\ \ifcase\month\or
  January\or February\or March\or April\or May\or June\or
  July\or August\or September\or October\or November\or
December\fi
  \space \number\year}
\def\monthyear{\ifcase\month\or
  January\or February\or March\or April\or May\or June\or
  July\or August\or September\or October\or November\or
December\fi
  \space \number\year}

%
\newcount\secno      
\newcount\subno      
\newcount\subsubno   
\newcount\appno      
\newcount\tableno    
\newcount\figureno   
%

%
\normalbaselineskip=20 truept
\baselineskip=20 truept

%
%
\def\title#1
   {\vglue1truein
   {\baselineskip=24 truept
    \pretolerance=10000
    \raggedright
    \noindent \fourteenpoint\bf #1\par}
    \vskip1truein minus36pt}
%

%
\def\author#1
  {{\pretolerance=10000
    \raggedright
    \noindent {\large #1}\par}}

%
\def\address#1
   {\bigskip
    \noindent \rm #1\par}

%
\def\shorttitle#1
   {\vfill
    \noindent \rm Short title: {\sl #1}\par
    \medskip}

%
\def\pacs#1
   {\noindent \rm PACS number(s): #1\par
    \medskip}

%
\def\jnl#1
   {\noindent \rm Submitted to: {\sl #1}\par
    \medskip}

%
\def\date
   {\noindent Date: \today\par
    \medskip}

%

%
\def\keyword#1
   {\bigskip
    \noindent {\bf Keyword abstract: }\rm#1}

%

%
%

%
\def\entry#1#2#3
   {\noindent
    \hangindent=20pt
    \hangafter=1
    \hbox to20pt{#1 \hss}#2\hfill #3\par}

%
\def\subentry#1#2#3
   {\noindent
    \hangindent=40pt
    \hangafter=1
    \hskip20pt\hbox to20pt{#1 \hss}#2\hfill #3\par}
\def\checkforsub{\futurelet\nexttok\decide}
\def\ssf{\relax}
\def\decide{\if\nexttok\ssf\let\endspace=\nospace
                \else\let\endspace=\extraspace\fi\endspace}
\def\nospace{\nobreak\par\nobreak}
%
%
\def\section#1{%
    \goodbreak
    \vskip24pt plus12pt minus12pt
    \nobreak
    \gdef\extraspace{\nobreak\bigskip\noindent\ignorespaces}%
    \noindent
    \subno=0 \subsubno=0
    \global\advance\secno by 1
    \noindent {\bf \the\secno. #1}\par\checkforsub}

%
\def\subsection#1{%
     \goodbreak
     \vskip24pt plus12pt minus6pt
     \nobreak
     \gdef\extraspace{\nobreak\medskip\noindent\ignorespaces}%
     \noindent
     \subsubno=0
     \global\advance\subno by 1
     \noindent {\sl \the\secno.\the\subno. #1\par}\checkforsub}

%
\def\subsubsection#1{%
     \goodbreak
     \vskip15pt plus6pt minus6pt
     \nobreak\noindent
     \global\advance\subsubno by 1
     \noindent {\sl \the\secno.\the\subno.\the\subsubno. #1}\null.
     \ignorespaces}

%
\def\appendix#1
   {\vskip0pt plus.1\vsize\penalty-250
    \vskip0pt plus-.1\vsize\vskip24pt plus12pt minus6pt
    \subno=0
    \global\advance\appno by 1
    \noindent {\bf Appendix \the\appno. #1\par}
    \bigskip
    \noindent}

%
\def\subappendix#1
   {\vskip-\lastskip
    \vskip36pt plus12pt minus12pt
    \bigbreak
    \global\advance\subno by 1
    \noindent {\sl \the\appno.\the\subno. #1\par}
    \nobreak
    \medskip
    \noindent}

%
\def\ack
   {\vskip-\lastskip
    \vskip36pt plus12pt minus12pt
    \bigbreak
    \noindent{\bf Acknowledgments\par}
    \nobreak
    \bigskip
    \noindent}


%

%
\def\tabcaption#1
   {\global\advance\tableno by 1
    \noindent {\bf Table \the\tableno.} \rm#1\par
    \bigskip}

%

%

%

%

%

%
\def\figcaption#1
   {\global\advance\figureno by 1
    \noindent {\bf Figure \the\figureno.} \rm#1\par
    \bigskip}

%

%

%
\def\refjl#1#2#3#4
   {\hangindent=16pt
    \hangafter=1
    \rm #1
   {\frenchspacing\sl #2
    \bf #3}
    #4\par}

%
\def\refbk#1#2#3
   {\hangindent=16pt
    \hangafter=1
    \rm #1
   {\frenchspacing\sl #2}
    #3\par}

%
\def\numrefjl#1#2#3#4#5
   {\parindent=40pt
    \hang
    \noindent
    \rm {\hbox to 30truept{\hss #1\quad}}#2
   {\frenchspacing\sl #3\/
    \bf #4}
    #5\par\parindent=16pt}

%
\def\numrefbk#1#2#3#4
   {\parindent=40pt
    \hang
    \noindent
    \rm {\hbox to 30truept{\hss #1\quad}}#2
   {\frenchspacing\sl #3\/}
    #4\par\parindent=16pt}

%

\def\ref#1{\par\noindent \hbox to 21pt{\hss
#1\quad}\frenchspacing\ignorespaces}

%
\def\frac#1#2{{#1 \over #2}}

%

%
\def\d{\hbox{\rm d}}

%
\def\e{\operatorname{e}}


\def\i{\operatorname{i}}
\chardef\ii="10

%

%

%

\catcode`\@=11
\def\vfootnote#1{\insert\footins\bgroup
    \interlinepenalty=\interfootnotelinepenalty
    \splittopskip=\ht\strutbox 
    \splitmaxdepth=\dp\strutbox \floatingpenalty=20000
    \leftskip=0pt \rightskip=0pt \spaceskip=0pt \xspaceskip=0pt
    \noindent\eightpoint\rm #1\ \ignorespaces\footstrut\futurelet\next\fo@t}

%
%
\def\eq(#1){\hfill\llap{(#1)}}
\catcode`\@=12
%
%



%
%





%
%

%
%

%
%

%

%

%

%
\def\gap{\;\lower3pt\hbox{$\buildrel > \over \sim$}\;}
%
%
\def\lap{\;\lower3pt\hbox{$\buildrel < \over \sim$}\;}
\def\tqs{\hbox to 25pt{\hfil}}


%

{\obeylines\gdef\startdisplay#1
  {\catcode`\^^M=5$$#1\halign\bgroup\indent##\hfil&&\qquad##\hfil\cr}}
\outer\def\enddisplay{\crcr\egroup$$}

\chardef\other=12
\def\ttverbatim{\begingroup \catcode`\\=\other \catcode`\{=\other
  \catcode`\}=\other \catcode`\$=\other \catcode`\&=\other
  \catcode`\#=\other \catcode`\%=\other \catcode`\~=\other
  \catcode`\_=\other \catcode`\^=\other
  \obeyspaces \obeylines \tt}
{\obeyspaces\gdef {\ }}  

\outer\def\begintt{$$\let\par=\endgraf \ttverbatim \parskip=0pt
  \catcode`\|=0 \rightskip=-5pc \ttfinish}
{\catcode`\|=0 |catcode`|\=\other 
  |obeylines 
  |gdef|ttfinish#1^^M#2\endtt{#1|vbox{#2}|endgroup$$}}

\catcode`\|=\active
{\obeylines\gdef|{\ttverbatim\spaceskip=.5em plus.25em minus.15em
                                            \let^^M=\ \let|=\endgroup}}%

\TagsOnRight

\tracingstats=1    

\font\twelverm=cmr10 scaled 1200

\baselineskip=15 truept
\normalbaselineskip=15pt

\def\CD{{\Cal D}}

\def\CM{{\Cal M}}

\def\ih{{\i\over\hbar}}
\def\overh{{1\over\hbar}}
\def\PSL{\operatorname{PSL}}
\def\SU{\operatorname{SU}}
\def\bbbr{\operatorname{{I\!R}}}                     
\def\bbbn{\operatorname{{I\!N}}}                     
\font\sans=cmssbx10
\def\sf{\sans}
\def\bbbm{\operatorname{{I\!M}}}
\def\bbbz{{\mathchoice {\hbox{$\sf\textstyle Z\kern-0.4em Z$}}
{\hbox{$\sf\textstyle Z\kern-0.4em Z$}}
{\hbox{$\sf\scriptstyle Z\kern-0.3em Z$}}
{\hbox{$\sf\scriptscriptstyle Z\kern-0.2em Z$}}}}    
\def\AA{{\hbox{\rm A}\NUM}}
\def\BB{{\hbox{\rm A}\NUM}}

\def\viert{1/4}
\def\half{{1\over2}}
\def\bhalf{\hbox{$\half$}}
\def\erfc{\operatorname{erfc}}
\def\myalign{\allowdisplaybreaks\align}
\def\dfrac{\dsize\frac}
\def\numcd{A.7}
\def\newenvironment#1#2#3#4{\long\def#1##1##2{\removelastskip
\vskip\baselineskip\noindent{#3#2\if!##1!.\else\unskip\ \ignorespaces
##1\unskip\fi\ }{#4\ignorespaces##2}\vskip\baselineskip}}
\newenvironment\lemma{Lemma}{\tenbf}{\it}
\newenvironment\proposition{Proposition}{\tenbf}{\it}
\newenvironment\theorem{Theorem}{\tenbf}{\it}
\newenvironment\corollary{Corollary}{\tenbf}{\it}
\def\vec#1{{\textfont1=\tenbf\scriptfont1=\sevenbf
\textfont0=\tenbf\scriptfont0=\sevenbf
\mathchoice{\hbox{$\displaystyle#1$}}{\hbox{$\textstyle#1$}}
{\hbox{$\scriptstyle#1$}}{\hbox{$\scriptscriptstyle#1$}}}}
\hfuzz=3pt

\newcount\Chapno
\def\PLUS{\advance\Chapno by 1}
\def\NUM{\the\Chapno}
\Chapno=0

\newcount\glno
\def\plus{\advance\glno by 1}
\def\minus{\advance\glno by -1}
\def\num{\the\glno}
\def\glnonull{\glno=0}

\newcount\Refno
\def\add{\advance\Refno by 1}
\Refno=1

\edef\ABD{\the\Refno}\add
\edef\AGHHa{\the\Refno}\add
\edef\AGHHb{\the\Refno}\add
\edef\AGSTA{\the\Refno}\add
\edef\BAVE{\the\Refno}\add
\edef\BVK{\the\Refno}\add
\edef\BAU{\the\Refno}\add
\edef\BF{\the\Refno}\add
\edef\BLWE{\the\Refno}\add
\edef\BOVO{\the\Refno}\add
\edef\BHR{\the\Refno}\add
\edef\BUGE{\the\Refno}\add
\edef\BUC{\the\Refno}\add
\edef\CIWI{\the\Refno}\add
\edef\CAR{\the\Refno}\add
\edef\CFG{\the\Refno}\add
\edef\CHc{\the\Refno}\add
\edef\CMS{\the\Refno}\add
\edef\DAGR{\the\Refno}\add
\edef\DURa{\the\Refno}\add
\edef\DK{\the\Refno}\add
\edef\EMOTb{\the\Refno}\add
\edef\FH{\the\Refno}\add
\edef\FLU{\the\Refno}\add
\edef\GASCH{\the\Refno}\add
\edef\GOOb{\the\Refno}\add
\edef\GODE{\the\Refno}\add
\edef\GOBR{\the\Refno}\add
\edef\GRA{\the\Refno}\add
\edef\GROb{\the\Refno}\add
\edef\GROe{\the\Refno}\add
\edef\GROh{\the\Refno}\add
\edef\GROm{\the\Refno}\add
\edef\GROs{\the\Refno}\add
\edef\GROu{\the\Refno}\add
\edef\GROw{\the\Refno}\add
\edef\GROx{\the\Refno}\add
\edef\GRSb{\the\Refno}\add
\edef\GRSf{\the\Refno}\add
\edef\GRSg{\the\Refno}\add
\edef\HEJ{\the\Refno}\add
\edef\HOI{\the\Refno}\add
\edef\HOST{\the\Refno}\add
\edef\INOb{\the\Refno}\add
\edef\JAMA{\the\Refno}\add
\edef\BJ{\the\Refno}\add
\edef\KLEh{\the\Refno}\add
\edef\KLEMUS{\the\Refno}\add
\edef\KOM{\the\Refno}\add
\edef\KUSRI{\the\Refno}\add
\edef\LABH{\the\Refno}\add
\edef\MANO{\the\Refno}\add
\edef\MERZ{\the\Refno}\add
\edef\PAKSa{\the\Refno}\add
\edef\PI{\the\Refno}\add
\edef\PT{\the\Refno}\add
\edef\PRAN{\the\Refno}\add
\edef\RS{\the\Refno}\add
\edef\RUND{\the\Refno}\add
\edef\SCATE{\the\Refno}\add
\edef\SCHUe{\the\Refno}\add
\edef\SEL{\the\Refno}\add
\edef\STEc{\the\Refno}\add
\edef\VEN{\the\Refno}\add


{\nopagenumbers
\pageno=0
\centerline{July 1993\hfill SISSA/119/93/FM}
\vskip1cm
\centerline{\fourteenpoint PATH INTEGRAL DISCUSSION}
\bigskip
\centerline{\fourteenpoint OF TWO AND- THREE-DIMENSIONAL}
\bigskip
\centerline{\fourteenpoint $\delta$-FUNCTION PERTURBATIONS}
\vskip1cm
\centerline{\twelverm CHRISTIAN GROSCHE$^{\hbox{*}}$}
\bigskip
\centerline{\it Scuola Internazionale Superiore di Studi Avanzati}
\centerline{\it International School for Advanced Studies}
\centerline{\it Via Beirut 4, 34014 Trieste, Italy}
\vfill\midinsert\narrower\noindent
{\bf Abstract.}
The incorporation of two- and three-dimensional $\delta$-function
perturbations into the path-integral formalism is discussed. In contrast
to the one-dimensional case, a regularization procedure is needed due to
the divergence of the Green-function $G^{(V)}(\vec x,\vec y;E)$, ($\vec
x,\vec y\in\bbbr^2,\bbbr^3$) for $\vec x=\vec y$, corresponding to a
potential problem $V(\vec x)$. The known procedure to define proper
self-adjoint extensions for Hamiltonians with deficiency indices can be
used to regularize the path integral, giving a perturbative approach
for $\delta$-function perturbations in two and three dimensions in the
context of path integrals. Several examples illustrate the formalism.
\endinsert
\bigskip
\centerline{\vrule height0.25pt depth0.25pt width4cm\hfill}
\noindent
{\sevenrm $^*$ Address from August 1993: II.Institut f\"ur Theoretische
          Physik, Universit\"at Hamburg, Luruper Chaussee 149,
          22761 Hamburg, Germany.}
\eject\pageno=0\centerline{\ }\vfill\eject}
\pageno=1


\glnonull\PLUS               
\section{Introduction}
$\delta$-function perturbations play an important r\^ole in quantum
physics because they can serve as simple models for more complicated
interactions and, more important, they are in many cases explicitly
solvable [\AGHHb]. They are also called point interactions (e.g.\ the
Fermi point interaction [\BLWE]) in nuclear physics, solid state
physics, and for the other fundamental interactions. The incorporation
of a $\delta$-function interaction in one-dimensional models causes no
serious problems, and is usually discussed as a simple solvable quantum
model in many textbooks in quantum mechanics (e.g.\ [\MERZ]).

The construction of the propagator can be done by a perturbative
approach in the path integral. Let $W(x)=V(x)-\gamma\delta(x-a)$
($a,x\in\bbbr$), and we consider the $\delta$-function as a perturbation
of the potential $V$. Then
\plus$$G^{(\delta)}(x'',x';E)
  =G^{(V)}(x'',x';E)+\big(\Gamma_{\gamma,a}^{(V)}(E)\big)^{-1}
   G^{(V)}(x'',a;E)G^{(V)}(a,x';E)
  \tag\NUM.\num$$
\edef\numaa{\NUM.\num}%
with $\Gamma_{\gamma,a}^{(V)}=1/\gamma-G^{(V)}(a,a;E)$, and $G^{(V)}
(E)$ and $G^{(\delta)}(E)$ are the Green functions (resolvent kernel)
of the unperturbed, respectively perturbed problem. The general
structure of (\numaa) is also known as Krein's formula [\AGHHb].

A comprehensive approach can be found in [\GROh], including numerous
examples, based on earlier work [\BAU, \GODE, \LABH], and, of course, on
the monograph of Albeverio et al.~[\AGHHb]. Considering the limit
$\gamma\to-\infty$, i.e.\ making the strength of the $\delta$-function
perturbation infinitely repulsive, gives a boundary value problem with
Dirichlet boundary-conditions at the boundary $x=a$, therefore
explicitly incorporating boundary problems in the path integral [\CAR,
\CFG, \CMS, \GROw, \GROx]. Note that for a symmetrical model, i.e.\
$V(x)=V(-x)$, a $\delta$-function perturbation at $x=0$ yields in the
limit $\gamma\to\infty$ a doubly generated energy level spectrum with
one ``spurious ground state'' $\tilde\Psi_0$ with infinite negative
energy and a corresponding wave-function concentrated at $x=0$, i.e.\
$\vert\tilde\Psi_0(x)\vert^2=\delta(x)$ (e.g.~[\AGSTA] and references
therein).

One-dimensional examples, however, describe only quantum mechanical
models on a line, respectively models which can be reduced to
one-dimensional models, e.g.\ radial problems. More realistic ones must
be at least two-, respectively three-dimensional. They can represent
lattices of interaction centers in thin films (e.g.~[\PRAN]) and solids,
respectively, and can be chosen arbitrarily located with arbitrary
strength.

Historically, point interactions were introduced in nuclear physics
starting in the 1930s. But this type of interactions appear also in many
other branches of physics, electromagnetic interactions with vector
potential, etc. Also very important, they model lattices in solid state
physics, yielding in the limit of infinitely many interaction centers
an electronic band structure. Often the notion of a $\delta$-function
perturbation is replaced by the notion of particular boundary-conditions
which the $\delta$-function perturbations actually describe, c.f.\
[\AGHHb] for a comprehensive bibliography on this subject.

The purpose of this paper is to incorporate such models into the path
integral formalism. As we will see, a regularization procedure will be
needed. The origin of the emerging divergence can be easily seen. A
perturbation expansion consideration yields an expression, where the
Green function $G^{(V)}(E)$ of the unperturbed problem, say for the free
particle Green function, must be evaluated at both arguments being
equal, i.e.\ $``G^{(V)}(\vec x,\vec x;E)"$. In two dimensions Green
functions generally diverge logarithmically in this case, i.e.\ $\lim
_{\vec x\to\vec y}G(\vec x,\vec y;E)\propto\ln\vert\vec x-\vec y\vert$,
whereas in three dimensions there is a simple pole, i.e.\ $\lim_{\vec x
\to\vec y}G(\vec x,\vec y;E)\propto\vert\vec x-\vec y\vert^{-1}$.
Consequently a regularization procedure is necessary. It consists
basically on a proper definition of the corresponding Friedrich
extension, making the perturbed Hamiltonian self-adjoint. What has to
be done is the following:  ``In physical terms, the coupling constant
$\lambda$ in the heuristic expression $-\Delta+\lambda\delta_y$ has to
be `renormalized' and turns out to be of the form $\lambda=\eta+\alpha
\eta^2$, with $\eta$ infinitesimal and $\alpha\in(-\infty,\infty]$.
[\AGHHb, p.3]'', a topic which was first discussed by Berezin and
Faddeev [\BF]. To put it into the language of approximating, the
$\delta$-function perturbation is rewritten in terms of a Gaussian
packet $\psi$ with parameter $\epsilon$, i.e.\ $\psi_\epsilon(x)=\e
^{-(x-a)^2/\epsilon}/\sqrt{\pi\epsilon}$, and for $D=3$, say, $\lambda$
in the expression $\lambda\psi_\epsilon(x)\psi_\epsilon(y)\psi_\epsilon
(z)$ has to be explicitly dependent on $\epsilon$, i.e.\ $\lambda=
\lambda(\epsilon)$. However, this is only necessary for $D=2,3$. For
$D=1$ no regularization is needed, and for $D\geq4$, the corresponding
Hamiltonian is already self-adjoint [\RS], c.f.\ the note in the
summary.

This kind of regularization procedure is known from functional analysis
[\AGHHb] to define the corresponding proper self-adjoint Hamiltonian,
and our purpose is to show that it can also be applied to regularize the
corresponding path integral formulation. We will find that the entire
Green function has to be taken into account in the regularization
procedure, and not only part of it (compare [\KUSRI]).

Such regularization procedures are well-known in other branches of
mathematical physics, let alone high energy physics, respectively
renormalization and quantum field theory. For instance, in quantum
mechanics let us note the Selberg trace formula [\GROu, \HEJ, \SEL,
\VEN], where for non-compact Riemann surfaces, the trace of a properly
chosen operator-valued function for the Laplacian on the non-compact
Riemann surface must be regulated by a subtraction of Eisenstein series
representing the continuous spectrum. This regularization does not
simply cancels the continuous part of the trace and na\"\ii vely gives
zero-terms in the trace formula, but instead leads to inadmissible
contributions, important in the theory of the Selberg trace formula,
and related fields, e.g.\ in number theory.

The further content of this paper will be as follows. In the next
section the theory of incorporating two- and three-dimensional
$\delta$-function perturbations into the path integral will be
presented. The regularization procedure will be applied for the
perturbation expansion. In the third section, several examples
illustrate the formalism, and will include the free particle, the
harmonic oscillator, and the Coulomb potential ($1/r$-) case on the
one hand. On the other, some models for nuclear physics potentials
will be given. The fourth sections concludes with a summary and a
discussion of the results. In appendix 1 the computation of the Green
function for the one-dimensional harmonic oscillator is sketched, and in
appendix 2 the evaluation of the propagator for a free particle with a
point interaction in two dimension will be given.
\goodbreak

\newpage
\glnonull\PLUS               
\section{Time ordered perturbation expansion of the path integral and
perturbation summation for $\delta$-function potentials}
The general method for the time-ordered perturbation expansion is
simple. We assume that we have a potential $W(\vec x)=V(\vec x)+\tilde
V(\vec x)$ ($\vec x\in\bbbr^D$) in the path integral, where it is
assumed that $W$ is so complicated that a  direct path integration is
not possible. However, the path integral corresponding to $V(\vec x)$ is
assumed to be known, which we call $K^{(V)}(T)$. We expand the path
integral containing $V(\vec x)$ in a perturbation expansion about
$\tilde V(\vec x)$ in the following way. The initial kernel
corresponding to $V$ propagates in $\epsilon$-time unperturbed, then it
is interacting with $\tilde V$, propagates again in another
$\epsilon$-time unperturbed, a.s.o, up to the final state. Let us
denote the path integral (Feynman kernel, respectively) corresponding to
the potential $V$ by ($T=t''-t'$)
\plus$$K^{(V)}(\vec x'',\vec x';T)
  =\int\limits_{\vec x(t')=\vec x'}^{\vec x(t'')=\vec x''}
  \CD^D\vec x(t)\exp\Bigg\{\ih\int_{t'}^{t''}
  \bigg[{m\over2}{\dot{\vec x}}^2-V(\vec x)\bigg]dt\Bigg\}\enspace.
  \tag\NUM.\num$$
We introduce the (energy-dependent) Green function (resolvent kernel)
\plus$$\aligned
  G^{(V)}(\vec x'',\vec x';E)
  &=\ih\int_0^\infty dT\,\e^{\i ET/\hbar}K^{(V)}(\vec x'',\vec x';T)
  \\
  K^{(V)}(\vec x'',\vec x';T)
  &={1\over2\pi\i}\int_{-\infty}^{\infty}
  G^{(V)}(\vec x'',\vec x';E)\e^{-\i ET\hbar}dE\enspace.
  \endaligned
  \tag\NUM.\num$$
\edef\numba{\NUM.\num}%
This gives the series expansion (see also e.g.\ [\BAU, \FH, \GODE,
\GROh, \LABH, \SCHUe])
\plus$$\myalign
  &K(\vec x'',\vec x';T)
         \\   &
  =\int\limits_{\vec x(t')=\vec x'}^{\vec x(t'')=\vec x''}\CD^D\vec x(t)
  \exp\Bigg\{\ih\int_{t'}^{t''}\bigg[
  {m\over2}{\dot{\vec x}}^2-V(\vec x)-\tilde V(\vec x)\bigg]dt\Bigg\}
         \\   &
  =K^{(V)}(\vec x'',\vec x';T)+\sum_{n=1}^\infty\bigg(-\ih\bigg)^n
  {1\over n!}\prod_{j=1}^n\int_{\bbbr^D}d^D\vec x_j
  \int_{t'}^{t''}dt_j
         \\   &\qquad\times
  \vphantom{\bigg]^{1/2}}
  K^{(V)}(\vec x_1,\vec x';t_1)\widetilde V(\vec x_1)
  K^{(V)}(\vec x_2,\vec x_1;t_2-t_1)
  \times\dotsc
         \\   &\qquad\dots\times
  \widetilde V(x_{n-1})K^{(V)}(\vec x_n,\vec x_{n-1};t_n-t_{n-1})
  \widetilde V(\vec x_n)K^{(V)}(\vec x'',\vec x_n;T-t_n)
  \vphantom{\bigg]^{1/2}}
         \\   &
  =K^{(V)}(\vec x'',\vec x';T)+\sum_{n=1}^\infty\bigg(-\ih\bigg)^n
  \prod_{j=1}^n\int_{t'}^{t_{j+1}} dt_j\int_{\bbbr^D}d^D\vec x_j
         \\   &\qquad\times
  \vphantom{\bigg]^{1/2}}
  K^{(V)}(\vec x_1,\vec x';t_1-t')
  \widetilde V(x_1)K^{(V)}(\vec x_2,\vec x_1;t_2-t_1)
  \times\dotsc
         \\   &\qquad\dots\times
  \widetilde V(x_{n-1})K^{(V)}(\vec x_n,\vec x_{n-1};t_n-t_{n-1})
  \widetilde V(\vec x_n)K^{(V)}(\vec x'',\vec x_n;t''-t_n)\enspace.
  \vphantom{\bigg]^{1/2}}
  \tag\NUM.\num\endalign$$
\edef\numbd{\NUM.\num}%
In the second step we have ordered the time as $t'=t_0<t_1<t_2<\dots<
t_{n+1}=t''$ and paid attention to the fact that $K^{(V)}(t_j-t_{j-1})$
is different from zero only if $t_j>t_{j-1}$.

We consider now an arbitrary potential $V(\vec x)$ in {\it one\/}
dimension with an additional $\delta$-function perturbation so that
[\GROh]
\plus$$W(x)=V(x)-\gamma\delta(x-a)\enspace.
  \tag\NUM.\num$$
The path integral for this potential problem reads
\plus$$K^{(W)}(x'',x';T)
  =\int\limits_{x(t')=x'}^{x(t'')=x''}\CD x(t)
   \exp\Bigg\{\ih\int_{t'}^{t''}\bigg[
   {m\over2}\dot x^2-W(x)\bigg]dt\Bigg\}\enspace.
  \tag\NUM.\num$$
Introducing the Green function $G(E)$ of the perturbed system similarly
to (\numba), and, for the time being, we assume that $G^{(V)}(a,a;E)$
exists. We then obtain due to the convolution theorem of the
Laplace-Fourier transformation
\plus$$G^{(\delta)}(x'',x';E)
  =G^{(V)}(x'',x';E)-
   \dsize\thickfrac{G^{(V)}(x'',a;E)G^{(V)}(a,x';E)}
              {G^{(V)}(a,a;E)-1/\gamma}\enspace.
  \tag\NUM.\num$$
\edef\numba{\NUM.\num}%
Let us consider $\delta$-function perturbations in {\it two\/} and {\it
three\/} dimensions. A formal series summation produces a formula
similar to (\numba). However, a simple example, say the free particle
Green function $G^{(0)}(\vec x'',\vec x';E)$, shows that the denominator
for $D>1$ does not exist. One must circumvent the arising divergence
and must regularize properly. One chooses the coupling ``$\gamma$''
appropriately, say, according to
\plus$${1\over\gamma}\mapsto \alpha+G^{(V)}(\vec a,\vec b;0)\enspace,
  \tag\NUM.\num$$
and considers the expression $G^{(V)}(\vec a,\vec b;0)-G^{(V)}(\vec a,
\vec b;E)$ in the limit $\vec b\to\vec a$, say. We see that in some
sense the coupling $\gamma$ has to be zero in a ``suitable way'' in
order to make the final expression well defined.

This formal reasoning can be put more rigorously [\AGHHb]. As a simple
example we start with the free Hamiltonian in $\bbbr^3$. It is
transformed via a momentum Fourier transformation into a multiplication
operator, so that $H=\hbar^2\vert\vec p\vert^2/2m$. A cut-off $\omega$
is introduced, and the coupling $\gamma$ is made explicitly dependent
on the cut-off, i.e.\ $\gamma=\gamma (\omega)$. All what remains is to
choose $\gamma(\omega)$ ``in such a way that the [perturbed operator
with cut-off $\omega$] $\hat H^\omega$ has a reasonable and nontrivial
limit [$\hat H$] as we remove the cut-off, i.e.\ as $\omega$ tends to
infinity.'' [\AGHHb, p.111]. For the free particle case we chose
\plus$${1\over\gamma(\omega)}=
  \alpha+{2m\over\hbar^2}{1\over(2\pi)^3}
  \int_{\vert\vec p\vert\leq\omega}{d^3\vec p\over\vert\vec p\vert^2}
  =\alpha+{m\over2\pi\hbar^2}\omega\enspace,
  \tag\NUM.\num$$
with $\alpha\in\bbbr$. We introduce the eigen-functions $\phi_{\vec y}
(\vec p)=(2\pi)^{-3/2}\e^{-\i \vec p\cdot\vec y}$ of the momentum
operator, and the corresponding cut-off eigen-functions $\phi_{\vec
y}^\omega= \chi_\omega\phi_{\vec y}$, where
\plus$$\chi_\omega(\vec p)=\left\{\alignedat 3
  &1   &\qquad   &\vert\vec p\vert\leq\omega\enspace,   \\
  &0   &\qquad   &\vert\vec p\vert>\omega   \enspace.
  \endalignedat\right.
  \tag\NUM.\num$$
Then ($E=\hbar^2\vert\vec k^2\vert/2m$)
\plus$$\myalign
  {1\over\gamma(\omega)}&
  -{2m\over\hbar^2} \bigg(\phi_{\vec y}^\omega,
   {1\over\vert\vec p\vert^2-\vert\vec k\vert^2}
   \phi_{\vec y}^\omega\bigg)
  \\   &
  =\alpha+{2m\over\hbar^2}{1\over(2\pi)^3}
  \int_{\vert\vec p\vert\leq\omega}d^3\vec p
  \bigg({1\over\vert\vec p\vert^2}
  -{1\over\vert\vec p\vert^2-\vert\vec k\vert^2}\bigg)
  \\   &
  \hbox{($\omega\to\infty$)}
  \\   &
  =\alpha+{m\over2\pi\hbar^3}\sqrt{-2mE}
  \enspace.
  \tag\NUM.\num\endalign$$
\edef\numbc{\NUM.\num}%
It follows that the perturbed cut-off operator $\hat H^\omega$ (in
$\vec p$-space) converges in norm resolvent sense to a self-adjoint
operator, respectively the perturbed cut-off operator $\e^{-\i\hat
H^\omega T/\hbar}$ to a unitary operator, hence the path integral
$K^{(\alpha)}(\vec x'',\vec x';T)=<\vec x''\vert \e^{-\i\hat H
T/\hbar}\vert\vec x'>$ exists and is well defined, with the
corresponding resolvent kernel given by [\AGHHb, Theorem 1.1.1, p.113]
\plus$$\myalign
  G^{(\alpha)}(\vec x'',\vec x';E)
  &
  =G_3^{(0)}(\vec x'',\vec x';E)
  +G_3^{(0)}(\vec x'',\vec a;E)G_3^{(0)}(\vec a,\vec x';E)
  \\   &\qquad\times
   \lim_{\omega\to\infty}
  \sum_{n=1}^\infty\bigg({\gamma(\omega)\over\hbar}\bigg)^n
  \bigg[{2m\over\hbar^2} \bigg(\phi_{\vec y}^\omega,
   {1\over\vert\vec p\vert^2-\vert\vec k\vert^2}
   \phi_{\vec y}^\omega\bigg)\bigg]^{n-1}
  \\   &
  =G_3^{(0)}(\vec x'',\vec x';E)
  +\lim_{\omega\to\infty}\dsize\thickfrac
   {G_3^{(0)}(\vec x'',\vec a;E)G_3^{(0)}(\vec a,\vec x';E)}
  {\dsize{1\over\gamma(\omega)}
  -\dsize{2m\over\hbar^2} \bigg(\phi_{\vec y}^\omega,
   \dsize{1\over\vert\vec p\vert^2-\vert\vec k\vert^2}
   \phi_{\vec y}^\omega\bigg)}
  \\   &
  =G_3^{(0)}(\vec x'',\vec x';E)
  +\dsize\frac
   { G_3^{(0)}(\vec x'',\vec a;E)G_3^{(0)}(\vec a,\vec x';E)}
   {\Gamma_{\alpha,\vec a}^{(0)}}
  \enspace,
  \tag\NUM.\num a\\
  \Gamma_{\alpha,\vec a}^{(0)}&=\alpha+{m\over2\pi\hbar^3}\sqrt{-2mE}
  \enspace,
  \tag\NUM.\num b\\
  G_3^{(0)}(x'',x';E)&={m\over4\pi\hbar^2\vert\vec x''-\vec x'\vert}
   \exp\bigg(-\dsize{\sqrt{-2mE}\over\hbar}
  \vert\vec x''-\vec x'\vert\bigg)\enspace.
  \tag\NUM.\num c\endalign$$
\edef\numbb{\NUM.\num}%
$G_3^{(0)}(E)$ denotes the free particle Green function in three
dimensions. Therefore we have obtained the three--dimensional
generalization of (\numba) for $V\equiv0$ and we conclude that the
perturbation expansion (\numbd) together with the regularization
(\numbc) is well defined and (\numbb) justifies our perturbation
expansion approach, respectively its formal series summation (\numba)
generalized to the three-dimensional case. This is in accordance  from
general considerations that the perturbed Green function for a
$\delta$-function perturbation located at $\vec x=\vec a$ ($\vec a\in
\bbbr^2,\bbbr^3$) must look like (\numbb) (Krein's formula [\AGHHb],
appendix A)

Let us note that the quantity $-1/4\pi\alpha$ represents the
scattering length of the perturbed operator.

However, an analogous line of reasoning is valid for the incorporation
of potentials $V(\vec x)$. Instead of performing the usual Fourier
transformation into $\vec p$-space, transforming the free Hamiltonian
into a multiplication operator, we perform a generalized Fourier
transformation (decomposition) in terms of its spectral expansion,
taking the corresponding eigen-functions $\phi_{E_\lambda}$ with energy
$E_\lambda$ and momentum $\vec p^2_\lambda=2mE_\lambda/\hbar^2$,
yielding a generalization of (\numbb), with $G^{(0)}(E)$ replaced by
$G^{(V)}(E)$.

Another way to determine $\Gamma_{\vec a,\gamma}^{(V)}(E)$, which also
includes potential problems in a more constructive way, exploits that
the Hilbert space of the perturbed problem must be properly defined,
i.e.\ one has to find the correct self-adjoint extension of the
heuristic operator ([\AGHHa, \AGHHb, \BUGE, \DAGR] and references
therein):
\plus$$H_\gamma=-{\hbar^2\over2m}\Delta+V(\vec x)
  +``\gamma\delta(\vec x-\vec a)"\enspace,
  \tag\NUM.\num$$
and $\Delta$ denotes the Laplace-Beltrami operator in two, respectively
three dimensions. Let $g(r)$ be a solution of the corresponding minimal
(reduced) radial $s$-wave Schr\"odinger operator ($\half\leq\lambda<
{3\over2}$, $\beta,\eta\in\bbbr$, $0<a<2$)
\plus$$\dot H=-{\hbar^2\over2m}{\d^2\over\d r^2}
     +{\hbar^2\over2m}{\lambda(\lambda-1)\over r^2}
     +{\eta\over r}+\beta r^{-a}+\hat V(r),\qquad
  D(\dot H)=C_0^\infty(\bbbr^+)\enspace,
  \tag\NUM.\num$$
\edef\numbe{\NUM.\num}%
where the notion $\dot H$ denotes the incorporation of a
$\delta$-function perturbation at $r=0$, say ($\dot H$ has deficiency
indices $(1,1)$), and $\hat V(r)\in L^\infty(\bbbr^+)$. We then can
formulate the following theorem:

\theorem{[\AGHHb, \BUGE]}{
Let
\plus$$F_\lambda^{(0)}(r)=r^\lambda\enspace,\qquad
  G_\lambda^{(0)}(r)=\left\{\alignedat 3
  -&{m\over\pi\hbar^2}\sqrt{r}\,\ln r  &\qquad  &\lambda=\half
                                       \enspace,  \\
  &{m\over2\pi\hbar^2} &\qquad  &\lambda=1     \enspace.
  \endalignedat\right.
  \tag\NUM.\num$$
All self-adjoint extensions of the operator $\dot H$ are given by
\plus$$\myalign
  H_\nu&=-{\hbar^2\over2m}{\d^2\over\d r^2}
     +{\hbar^2\over2m}{\lambda(\lambda-1)\over r^2}
     +{\eta\over r}+\beta r^{-a}+\hat V(r)\enspace,
  \tag\NUM.\num a\\
  \CD(H_\nu)&=\bigg\{
     g\in L^2(\bbbr^+)\bigg\vert g,g'\in AC_{loc}(\bbbr^+);
     \alpha g_{0,\lambda}=g_{1,\lambda};H_\nu g\in L^2(\bbbr^+)\bigg\}
  \\   &
  \tag\NUM.\num b\\
  -\infty&<\alpha\leq\infty,\quad\half\leq\lambda<{3\over2},\quad
    \beta,\eta\in\bbbr,\qquad0<a<2\enspace.
  \tag\NUM.\num c\endalign$$
\edef\numbh{\NUM.\num}%
$\hat V\in L^\infty(\bbbr^+)$ real valued, and $AC_{loc}(\bbbm)$ denotes
the set of absolutely continuous functions on $\bbbm$ (here $=\bbbr^+$).
The boundary values $g_{0,\lambda}$, $g_{1,\lambda}$ are defined by
\plus$$g_{0,\lambda}=\lim_{r\to0^+}{g(r)\over G_\lambda^{(0)}(r)}
  ,\qquad
  g_{1,\lambda}=\lim_{r\to0^+}{g(r)-g_{0,\lambda}G_\lambda^{B}(r)
                                \over F_\lambda^{(0)}(r)}\enspace,
  \tag\NUM.\num$$
where $G_\lambda^B(r)$ denotes the asymptotic expansion of the
irregular solution $G_\lambda(r)$ of (\numbh) up to order $r^t$,
$t\leq2\lambda-1$.
}
\plus\plus\edef\numbh{\NUM.\num}\plus%

Generally one has $G_\lambda^B(r)=G_\lambda^{(0)}(r)+\hbox{additional
terms}$. Two special cases of $G_\lambda^B(r)$ can be e.g.\ stated for
the $\lambda=\half,1$, i.e.\ for the Schr\"odinger operator in two and
three dimensions, respectively, and all other potential terms in
(\numbh) equal to zero, which will be sufficient for our purposes. Then
$$\myalign
   G_\half^B(r)&=G_\half^{(0)}(r)=-{m\over\pi\hbar^2}\sqrt{r}\,\ln r
  \enspace,
  \tag\NUM.\num\\   \global\plus
   G_1^B(r)&={m\over2\pi\hbar^2}
   \bigg(1-{m\eta\over\hbar^2}r-{2m\eta\over\hbar^2}r\ln r\bigg)
  \enspace.
  \tag\NUM.\num\endalign$$
Take into account the $\beta\not=0$ contributions considerably
complicates the expressions and will not be stated here, c.f~[\BUGE].

We consider $\lambda=(D-1)/2$ and usually we identify $F_\lambda\equiv
F_D$, $G_\lambda\equiv G_D$, etc.\ in the following. Note that the
additional potential $\hat V(r)$ does not play any further r\^ole in the
regularization. The condition $\half\leq\lambda<{3\over2}$ can be
physically interpreted as follows: Set $N=\lambda-\half$, then $\lambda
(\lambda-1)=N^2-1/4$ and the functional weight in the radial path
integral corresponding to the Hamiltonian (\numbh) [\GRSb] is
proportional to $\mu_N[r^2]$. The number $N$ corresponds to the
magnetic flux inside an infinitely thin Aharonov-Bohm solenoid. Only
fluxes less then unity are allowed and there exists exactly {\sl one\/}
bound state solution and the parameter of the self-adjoint extension
describes the anomaly of the magnetic moment of the particle bound by
the $\delta$-function-like solenoid [\BOVO].

Summarizing, we can state that the result of the regularization
procedure (\numbh) of the formal series summation (\numba) of the path
integral perturbation expansion (\numbd) with the heuristic expression
``$\gamma\delta(\vec x-\vec a)$'' in the Lagrangian in the path
integral in two and three dimensions yields in an analogous way as the
consideration in (\numbb) the Green function
\plus$$\myalign
  &\ih\int_0^\infty dT\,\e^{\i ET/\hbar}\left.
  \int\limits_{\vec x(t')=\vec x'}^{\vec x(t'')=\vec x''}\CD^D\vec x(t)
 \right\vert_{\scriptstyle 1/\gamma-\lim_{\vec x,\vec y\to\vec a}
          G^{(0)}(\vec x,\vec y;E)\hfill
          \atop\scriptstyle\mapsto
          \alpha g_{0,D}-g_{1,D}\hfill}
  \\   &\qquad\qquad\qquad\times
  \exp\Bigg\{\ih\int_{t'}^{t''}\bigg[{m\over2}{\dot{\vec x}}^2
  -V(\vec x)+\gamma\delta(\vec x-\vec a)\bigg]dt\Bigg\}
  \\   &
  \equiv\ih\int_0^\infty dT\,\e^{\i ET/\hbar}
  \int\limits_{\vec x(t')=\vec x'}^{\vec x(t'')=\vec x''}
  \CD_{\Gamma^{(V)}_{\alpha,\vec a}}^D\vec x(t)
  \exp\Bigg\{\ih\int_{t'}^{t''}\bigg[{m\over2}{\dot{\vec x}}^2
  -V(\vec x)+\gamma\delta(\vec x-\vec a)\bigg]dt\Bigg\}
  \\   &
  =:G^{(V)}(\vec x'',\vec x';E)
  +\dsize\thickfrac
   {G_3^{(V)}(\vec x'',\vec a;E)G_3^{(V)}(\vec a,\vec x';E)}
   {\Gamma_{\alpha,\vec a}^{(V)}}\enspace.
  \tag\NUM.\num\endalign$$
\edef\numbf{\NUM.\num}%
with $\Gamma^{(V)}_{\alpha,\vec a}$ given by
\plus$$\Gamma^{(V)}_{\alpha,\vec a}=\alpha g_{0,D}-g_{1,D}\enspace.
  \tag\NUM.\num$$
\edef\numbg{\NUM.\num}%
and we {\sl define\/} the heuristic point-interaction potential
$\gamma\delta(\vec x-\vec a)$ in the path integral by means of
(\numbf,\numbg) with the regularization prescription (\numbh).

A convenient way to find the function $g(r)$ is to set $g(r)=\Omega^{-1}
(D) G_{l=0}^{(V),red.}(r,r;E)$ $= \Omega^{-1}(D) r^\lambda G_{l=0}^{(V)}
(r,r;E)$, where $V$ includes all potential terms in (\numbe). Here
$\Omega(D)$ denotes the volume of the unit sphere $S^{(D-1)}$.

In three dimensions it is also possible to state the regularization
rule as follows [\AGHHb]
\plus$$\myalign
  &\ih\int_0^\infty dT\,\e^{\i ET/\hbar}
  \int\limits_{\vec x(t')=\vec x'}^{\vec x(t'')=\vec x''}
  \CD_{\Gamma^{(V)}_{\alpha,\vec a}}^D\vec x(t)
  \exp\Bigg\{\ih\int_{t'}^{t''}\bigg[{m\over2}{\dot{\vec x}}^2
  -V(\vec x)+\gamma\delta(\vec x-\vec a)\bigg]dt\Bigg\}
  \\   &
  =G^{(V)}(\vec x'',\vec x';E)
   +\big(\Gamma^{(V)}_{\alpha,\vec a}(E)\big)^{-1}
  G^{(V)}(\vec x'',\vec a;E)G^{(V)}(\vec a,\vec x';E)\enspace,
  \tag\NUM.\num\endalign$$
with $\Gamma^{(V)}_{\alpha,\vec a}$ given by
\plus$$\Gamma^{(V)}_{\alpha,\vec a}
   =\alpha+{\partial\over\partial r_{12}}
    r_{12}G^{(V)}(\vec x'',\vec x';E)
    \Bigg\vert_{\scriptstyle
                r_{12}=\vert\vec x''-\vec x'\vert=0
                \atop\scriptstyle
                \vec x',\vec x''=\vec a\hfill}\enspace,
  \tag\NUM.\num$$
provided an explicit expression for $G^{(V)}(\vec x'',\vec x';E)$
exists.
\goodbreak

\glnonull\PLUS               
\section{Examples}\nobreak\ssf
\subsection{One-dimensional examples}
For completeness we refer some one-dimensional examples. This includes
the free particle, the harmonic oscillator and the Coulomb potential.
No new material is presented. The results will give an easy possibility
to compare the results in one, two and three dimensions with each other,
respectively to introduce some notation. We also introduce the Green
functions for the free particle, the harmonic oscillator, and the
Coulomb potential. Of course, no regularization procedure is needed.

Let us note that an implicit equation for the time-dependent propagator
corresponding to (\numaa) is due to Gaveau and Schulman [\GASCH].
They obtained
\plus$$K^{(\delta)}(x'',x';T)=K^{(V)}(x'',x';T)+\i{\gamma\over\hbar}
  \int_{t'}^{t''}K^{(V)}(x'',a;t)K^{(\delta)}(a,x';T-t)dt\enspace.
  \tag\NUM.\num$$
\goodbreak

\subsubsection{The free particle}
Let us start with the free particle (labeled by ``$(0)$''). The path
integral of the $D$-dimensional free particle reads as ($\vec x\in
\bbbr^D$)
$$\myalign
  K^{(0)}(\vec x'',\vec x';T)&=
  \int\limits_{\vec x(t')=\vec x'}^{\vec x(t'')=\vec x''}\CD^D\vec x(t)
  \exp\left({\i m\over2\hbar}\int_{t'}^{t''}{\dot{\vec x}}^2dt\right)
  \\
  &=\bigg({m\over2\pi\i\hbar T}\bigg)^{D/2}
   \exp\bigg[-{m\over2\i\hbar T}(\vec x''-\vec x')^2\bigg]
  \tag\NUM.\num\\   \global\plus
  &={1\over(2\pi)^D}\int_{\bbbr^D} d^D\vec p
   \exp\bigg[\i\vec p\cdot(\vec x''-\vec x')
   -\i T{\hbar\vert\vec p\vert^2\over2m}\bigg]\enspace.
  \tag\NUM.\num\endalign$$
The energy dependent kernel is given by
\plus$$\myalign
  G^{(0)}_D&(\vec x'',\vec x';E)=
   \ih\int_0^\infty  dT\,\e^{\i TE/\hbar}
  \int\limits_{\vec x(t')=\vec x'}^{\vec x(t'')=\vec x''}\CD^D\vec x(t)
  \exp\left({\i m\over2\hbar}\int_{t'}^{t''}{\dot{\vec x}}^2dt\right)
  \\
  &=2\ih\bigg({m\over2\pi\i\hbar}\bigg)^{D/2}
  \bigg({m\over2E}\vert\vec x''-\vec x'\vert^2\bigg)^{\half(1-{D/2})}
   K_{1-{D\over2}}\bigg({\vert\vec x''-\vec x'\vert\over\hbar}
  \sqrt{-2mE}\bigg)\quad
  \\
  &=\ih\bigg({m\over2\hbar}\bigg)^{D/2}
   \bigg({m(-\pi)^2\over2E}
    \vert\vec x''-\vec x'\vert^2\bigg)^{\half(1-{D/2})}
   H_{1-{D\over2}}^{(1)}
  \bigg(\i{\vert\vec x''-\vec x'\vert\over\hbar}\sqrt{-2mE}\bigg)
  \enspace.
  \\   &
  \tag\NUM.\num\endalign$$
\edef\numce{\NUM.\num}%
For $D=1$ one has
\plus$$G^{(0)}_1(\vec x'',\vec x';E)
  ={1\over\hbar}\sqrt{-{m\over2E}}
  \exp\bigg(-{\vert x''-x'\vert\over\hbar}
        \sqrt{-2mE}\bigg)\enspace.
  \tag\NUM.\num$$
{}From the general theory we now have [\BAU, \GODE, \GROh, \LABH, \MANO]
\plus$$\myalign
  &\ih \int_0^\infty  dT\,\e^{\i TE/\hbar}
   \int\limits_{x(t')=x'}^{x(t'')=x''}\CD x(t)
  \exp\left\{\ih\int_{t'}^{t''}
  \bigg[{m\over2}\dot x^2+\gamma\delta(x-a)\bigg]dt\right\}
         \\   &
  ={1\over\hbar}\sqrt{-{m\over2E}}
  \exp\bigg(-{\sqrt{-2mE}\over\hbar}\,\vert x''-x'\vert\bigg)
        \\   &\qquad\qquad
  +{m\gamma\over2\hbar^2}\dsize\thickfrac
  {\exp\bigg[-\dsize\frac{\sqrt{-2mE}}{\hbar}
   (\vert x''-a\vert+\vert a-x'\vert)\bigg]}
  {\sqrt{-E}\bigg(\sqrt{-E}-\dsize\frac{\gamma}{\hbar}
  \sqrt{\dsize\frac{m}{2}}\bigg)}\enspace.
  \tag\NUM.\num\endalign$$
\edef\numcf{\NUM.\num}%
To determine the propagator from (\numcf) the inverse Laplace-Fourier
transformation must be applied, which can be explicitly done with
result [\GODE]
\plus$$\myalign
  &\int\limits_{x(t')=x'}^{x(t'')=x''}\CD x(t)
  \exp\left\{\ih\int_{t'}^{t''}
  \bigg[{m\over2}\dot x^2+\gamma\delta(x-a)\bigg]dt\right\}
         \\   &=
  \sqrt{m\over2\pi\i\hbar T}\,\exp\bigg[{\i m\over2\hbar T}
       (x''-x')^2\bigg]
         \\   &\qquad
  +{m\gamma\over2\hbar^2}
  \exp\bigg(-{m\gamma\over\hbar^2}(\vert x''-a\vert+\vert x'-a\vert)
         +\ih{m\gamma^2\over2\hbar}T\bigg)
         \\   &\qquad\qquad\times
  \erfc\Bigg[\sqrt{m\over2\i\hbar T}
  \bigg(\vert x''-a\vert+\vert x'-a\vert-\ih\gamma T\bigg)\Bigg]
  \enspace.
  \tag\NUM.\num\endalign$$
\goodbreak

\subsubsection{The harmonic oscillator}
The harmonic oscillator (labeled by ``$(\omega)$'') path integral is
given by [\FH]
\plus$$\myalign
  &\int\limits_{x(t')=x'}^{x(t'')=x''}\CD x(t)
   \exp\left[{\i m\over2\hbar}
   \int_{t'}^{t''}(\dot x^2-\omega^2x^2)dt\right]
  \\
  &=\bigg({m\omega\over2\pi\i\hbar\sin\omega T}\bigg)^{1/2}
   \exp\bigg\{-{m\omega\over2\i\hbar}
    \bigg[({x'}^2+{x''}^2)\cot\omega T
   -2{x'x''\over\sin\omega T}\bigg]\bigg\}
  \enspace.
  \tag\NUM.\num\endalign$$
\edef\numci{\NUM.\num}%
We do not discuss the case of caustics etc.
The energy dependent kernel can be evaluated to be given by
(c.f. [\BAVE], and appendix 1)
\plus$$\myalign
  & \ih \int_0^\infty  dT\,\e^{\i TE/\hbar}
  \int\limits_{x(t')=x'}^{x(t'')=x''}\CD x(t)
  \exp\left[{\i m\over2\hbar}
  \int_{t'}^{t''}(\dot x^2-\omega^2x^2)dt\right]
         \\   &
  =\sqrt{m\over\pi\hbar^3\omega}\,
     \Gamma\bigg(\half-{E\over\hbar\omega}\bigg)
  D_{-\half+{E\over\hbar\omega}}\left(\sqrt{2m\omega\over\hbar}\,
   x_>\right)
  D_{-\half+{E\over\hbar\omega}}\left(-\sqrt{2m\omega\over\hbar}\,
   x_<\right)\enspace.
         \\   &
  \tag\NUM.\num\endalign$$
\edef\numcj{\NUM.\num}%
The general theory then gives for a perturbed harmonic oscillator
[\GROh] ($x''\geq a\geq x'$):
\plus$$\myalign
  & \ih \int_0^\infty  dT\,\e^{\i TE/\hbar}
  \int\limits_{x(t')=x'}^{x(t'')=x''}\CD x(t)
  \exp\left\{\ih\int_{t'}^{t''}
  \bigg[{m\over2}\dot x^2-{m\over2}\omega^2x^2
                         +\gamma\delta(x-a)\bigg]dt\right\}
         \\   &
   =\sqrt{m\over\pi\hbar^3\omega}
   \Gamma\bigg(\half-{E\over\hbar\omega}\bigg)
    D_{-\half+{E\over\hbar\omega}}
      \bigg(-\sqrt{2m\omega\over\hbar}\,x'\bigg)
    D_{-\half+{E\over\hbar\omega}}
     \bigg(\sqrt{2m\omega\over\hbar}\,x''\bigg)
         \\   &\qquad\times
   \Bigg[1-{\gamma\over\hbar}\sqrt{m\over\pi\hbar\omega}\,
   \Gamma\bigg(\half-{E\over\hbar\omega}\bigg)
         \\   &\qquad\qquad\qquad\times
   D_{-\half+{E\over\hbar\omega}}
     \bigg(-\sqrt{2m\omega\over\hbar}\,a\bigg)
   D_{-\half+{E\over\hbar\omega}}
     \bigg(\sqrt{2m\omega\over\hbar}\,a\bigg)\Bigg]^{-1}\enspace,
  \tag\NUM.\num\endalign$$
($x''\geq x'\geq a$):
\plus$$\myalign
  & \ih \int_0^\infty  dT\,\e^{\i TE/\hbar}
  \int\limits_{x(t')=x'}^{x(t'')=x''}\CD x(t)
  \exp\left\{\ih\int_{t'}^{t''}
  \bigg[{m\over2}\dot x^2-{m\over2}\omega^2x^2
                         +\gamma\delta(x-a)\bigg]dt\right\}
         \\   &
   =\sqrt{m\over\pi\hbar^3\omega}\,
     \Gamma\bigg(\half-{E\over\hbar\omega}\bigg)
   D_{-\half+{E\over\hbar\omega}}
   \bigg(-\sqrt{2m\omega\over\hbar}\,x'\bigg)
   D_{-\half+{E\over\hbar\omega}}
   \bigg(\sqrt{2m\omega\over\hbar}\,x''\bigg)
         \\   &
   +{m\gamma\over\pi\omega\hbar^3}
   D_{-\half+{E\over\hbar\omega}}^2
     \bigg(-\sqrt{2m\omega\over\hbar}\,a\bigg)
   D_{-\half+{E\over\hbar\omega}}
     \bigg(\sqrt{2m\omega\over\hbar}\,x''\bigg)
   D_{-\half+{E\over\hbar\omega}}
     \bigg(\sqrt{2m\omega\over\hbar}\,x'\bigg)
         \\   &\qquad\times
   \Bigg[1-{\gamma\over\hbar}\sqrt{m\over\pi\hbar\omega}\,
   \Gamma\bigg(\half-{E\over\hbar\omega}\bigg)
         \\   &\qquad\qquad\qquad\times
   D_{-\half+{E\over\hbar\omega}}
     \bigg(-\sqrt{2m\omega\over\hbar}\,a\bigg)
   D_{-\half+{E\over\hbar\omega}}
     \bigg(\sqrt{2m\omega\over\hbar}\,a\bigg)\Bigg]^{-1}\enspace,
  \tag\NUM.\num\endalign$$
\edef\numcb{\NUM.\num}%
($x''\leq x'\leq a$):
\plus$$\myalign
  & \ih \int_0^\infty  dT\,\e^{\i TE/\hbar}
  \int\limits_{x(t')=x'}^{x(t'')=x''}\CD x(t)
  \exp\left\{\ih\int_{t'}^{t''}
  \bigg[{m\over2}\dot x^2-{m\over2}\omega^2x^2
                         +\gamma\delta(x-a)\bigg]dt\right\}
         \\   &
   =\sqrt{m\over\pi\hbar\omega}\,
     \Gamma\bigg(\half-{E\over\hbar\omega}\bigg)
   D_{-\half+{E\over\hbar\omega}}
  \bigg(-\sqrt{2m\omega\over\hbar}\,x''\bigg)
   D_{-\half+{E\over\hbar\omega}}
  \bigg(\sqrt{m\omega\over2\hbar}\,x'\bigg)
         \\   &
   +{m\gamma\over\pi\omega\hbar^3}
   D_{-\half+{E\over\hbar\omega}}^2
     \bigg(\sqrt{2m\omega\over\hbar}\,a\bigg)
   D_{-\half+{E\over\hbar\omega}}
     \bigg(-\sqrt{2m\omega\over\hbar}\,x'\bigg)
   D_{-\half+{E\over\hbar\omega}}
     \bigg(-\sqrt{2m\omega\over\hbar}\,x''\bigg)
         \\   &\qquad\times
   \Bigg[1-{\gamma\over\hbar}\sqrt{m\over\pi\hbar\omega}\,
   \Gamma\bigg(\half-{E\over\hbar\omega}\bigg)
         \\   &\qquad\qquad\qquad\times
   D_{-\half+{E\over\hbar\omega}}
     \bigg(-\sqrt{2m\omega\over\hbar}\,a\bigg)
   D_{-\half+{E\over\hbar\omega}}
     \bigg(\sqrt{2m\omega\over\hbar}\,a\bigg)\Bigg]^{-1}\enspace.
  \tag\NUM.\num\endalign$$
\edef\numcc{\NUM.\num}%
\goodbreak

\subsubsection{The Coulomb potential}
The Coulomb potential (labeled by ``$(C)$'') in one dimension is also
know as the Kratzer potential. Its corresponding Green function via
path integration can be done by considering the $D$-dimensional Coulomb
potential, say, and then restricting to $D=1$ [\CHc]. Of course, the
problem is closely related to a Kustaanheimo-Stiefel transformation in
the in the path integral, hence to the space-time transformation
technique in path integrals. This has been discussed by many authors,
let us mention [\CHc, \DK, \GROm, \GRSb, \HOI, \INOb, \KLEh, \PAKSa,
\STEc] and references therein. A comprehensive survey will be given in
[\GRSg].

We now have ($x>0$, $\kappa=q_1q_2\sqrt{-m/2E}/\hbar$)
\plus$$\myalign
  & \ih \int_0^\infty  dT\,\e^{\i TE/\hbar}
  \int\limits_{x(t')=x'}^{x(t'')=x''}\CD x(t)\exp\left[\ih
  \int_{t'}^{t''}\bigg(
  {m\over2}\dot x^2+{q_1q_2\over\vert x\vert}\bigg)dt\right]
         \\   &
  ={1\over\hbar}\sqrt{-{m\over2E}}\Gamma(1-\kappa)
     W_{\kappa,1/2}\bigg(\sqrt{-8mE}\,{x_>\over\hbar}\bigg)
     M_{\kappa,1/2}\bigg(\sqrt{-8mE}\,{x_<\over\hbar}\bigg)
  \enspace.
  \tag\NUM.\num\endalign$$
For a $\delta$-perturbed $1/r$-potential we then obtain [\GROh]
($\kappa=q_1q_2\sqrt{-m/2E}\,/\hbar$, $x''\geq a\geq x'$)
\plus$$\myalign
  & \ih \int_0^\infty  dT\,\e^{\i TE/\hbar}
  \int\limits_{x(t')=x'}^{x(t'')=x''}\CD x(t)
  \exp\left\{\ih\int_{t'}^{t''}
  \bigg[{m\over2}\dot x^2+{q^2\over x}+\gamma\delta(x-a)\bigg]dt\right\}
         \\   &
  =\overh\sqrt{-{m\over2E}}\,\Gamma(1-\kappa)
  W_{\kappa,\half}\bigg({x''\over\hbar}\sqrt{-8mE}\,\bigg)
  M_{\kappa,\half}\bigg({x'\over\hbar}\sqrt{-8mE}\,\bigg)
         \\   &\qquad\times
  \Bigg[1-{\gamma\over\hbar^2}\sqrt{-{m\over2E}}\,\Gamma(1-\kappa)
  W_{\kappa,\half}\bigg({a\over\hbar}\sqrt{-8mE}\,\bigg)
  M_{\kappa,\half}\bigg({a\over\hbar}\sqrt{-8mE}\,\bigg)\Bigg]^{-1}
         \\   &
  \tag\NUM.\num\endalign$$
and similarly as in (\numcb,\numcc) for the other cases.
\goodbreak

\subsubsection{Multiple $\delta$-function perturbations}
For completeness we also cite the case of a multiple
$\delta$-function perturbation. Since only the Green function $G^{(V)}
(E)$ of the unperturbed problem is relevant, the incorporation of
multiple $\delta$-function perturbations can be successively done with
result [\AGHHb, \GOBR, \GROw]
$$\myalign
  & \ih \int_0^\infty  dT\,\e^{\i TE/\hbar}
  \int\limits_{x(t')=x'}^{x(t'')=x''}\CD x(t)\exp\left\{\ih
   \int_{t'}^{t''}\left[{m\over2}\dot x^2-V(x)
   +\sum_{j=1}^N\gamma_j\delta(x-a_j)\right]dt\right\}
         \\   &
  =\dsize\thickfrac{\left\vert\matrix
  G^{(V)}(x'',x';E)  &G^{(V)}(x'',a_1;E) &\hdots
                                 &G^{(V)}(x'',a_N;E)         \\
  G^{(V)}(a_1,x';E)  &G^{(V)}(a_1,a_1;E)-1/\gamma_1
                 &\hdots         &G^{(V)}(a_1,a_N;E)         \\
  \vdots         &\vdots         &\ddots            &\vdots  \\
  G^{(V)}(a_N,x';E)  &G^{(V)}(a_N,a_1)   &\hdots
                                 &G^{(V)}(a_N,a_N;E)-1/\gamma_N
  \endmatrix\right\vert}{\left\vert\matrix
  G^{(V)}(a_1,a_1;E)-1/\gamma_1
                 &\hdots         &G^{(V)}(a_1,a_N;E)    \\
  \vdots         &\ddots         &\vdots         \\
  G^{(V)}(a_N,a_1;E) &\hdots
                 &G^{(V)}(a_N,a_N;E)-1/\gamma_N
  \endmatrix\right\vert}
  \tag\NUM.\num\\   \global\plus
  &=G^{(V)}(x'',x';E)
  -\sum_{j,j'=1}^N\Big(\Gamma^{(V)}_{\{\alpha\},\vec a}\Big)_{j,j'}^{-1}
  G^{(V)}(x'',a_j;E)G^{(V)}(a_{j'},x';E)\enspace,
  \tag\NUM.\num\endalign$$
with the matrix $\Gamma^{(V)}_{\{\alpha\},\{a\}}$ given by $\big(\{
\alpha\}=\{\gamma_k\}_{k=1}^N$, $\{a\}=\{a_k\}_{k=1}^N\big)$
\plus$$\big(\Gamma^{(V)}_{\{\alpha\},\{a\}}\big)_{j,j'}
  =G^{(V)}(a_j,a_{j'};E)-{\delta_{jj'}\over\gamma_j}\enspace.
  \tag\NUM.\num$$
The determinant and matrix representations are checked by induction
[\GROw].
\goodbreak

\newpage\noindent
\subsection{Two-dimensional examples}\nobreak\ssf
\subsubsection{The Free Particle}
We keep the ordering schema of the first subsection and start with the
free particle case. Form the general $D$-dimensional free particle
Green function we consider $D=2$ and obtain
\plus$$G_2^{(0)}(\vec x'',\vec x';E)={m\over\pi\hbar^2}
  K_0\bigg(\dsize{\sqrt{-2mE}\over\hbar}\vert\vec x''-\vec x'\vert\bigg)
  \enspace,
  \tag\NUM.\num$$
which is logarithmically divergent for $\vert\vec x''-\vec x'\vert\to0$
due to $K_0(z)\propto -\ln(z/2)+\Psi(1)$ ($z\to0$). Here $-\Psi(1)=
0.57721\,56649\, 01532\,86061\dots$ denotes Euler's constant. In order
to regularize the problem we apply the theory of section 2. We set
($r>0$)
\plus$$g(r)={m\over\pi\hbar^2}\sqrt{r}\,
  K_0\bigg(\dsize{\sqrt{-2mE}\over\hbar}r\bigg)\enspace,
  \tag\NUM.\num$$
from which follows
\plus$$g_{0,2}={m\over\pi\hbar^2}\enspace,\qquad
  g_{1,2}=\Psi(1)-\ln{\sqrt{-2mE}\over2\hbar}
  \tag\NUM.\num$$
($\vec x\in\bbbr^2 \setminus\{0\}$). We therefore obtain (compare also
[\AGHHa, \AGHHb])
$$\myalign
  &\ih\int_0^\infty dT\,\e^{\i ET/\hbar}
  \int\limits_{\vec x(t')=\vec x'}^{\vec x(t'')=\vec x''}
  \CD_{\Gamma^{(0)}_{\alpha,\vec a}}^2\vec x(t)
  \exp\Bigg\{\ih\int_{t'}^{t''}\bigg[{m\over2}{\dot{\vec x}}^2
  +\gamma\delta(\vec x-\vec a)\bigg]dt\Bigg\}
  \\   &
  ={m\over\pi\hbar^2}K_0\bigg({\sqrt{-2mE}\over\hbar}
         \vert\vec x''-\vec x'\vert\bigg)
  \\   &\qquad
  +\bigg({m\over\pi\hbar^2}\bigg)^2\dsize\thickfrac{
  K_0\bigg(\dsize{\sqrt{-2mE}\over\hbar}\vert\vec x''-\vec a\vert\bigg)
  K_0\bigg(\dsize{\sqrt{-2mE}\over\hbar}\vert\vec a-\vec x'\vert\bigg)
  } {\alpha+\dsize{m\over\pi\hbar^2}\bigg[
  \ln\dsize{\sqrt{-2mE}\over2\hbar}-\Psi(1)\bigg]}
  \tag\NUM.\num\\   \global\plus
        &
  ={m\over\pi\hbar^2}
  K_0\bigg(\sqrt{-2mE}\,{r_>\over\hbar}\bigg)
  I_0\bigg(\sqrt{-2mE}\,{r_<\over\hbar}\bigg)
  \\   &\qquad
  +\bigg({m\over\pi\hbar^2}\bigg)^2
  \dsize\thickfrac{
  K_0\bigg(\sqrt{-2mE}\,\dsize{r'\over\hbar}\bigg)
  K_0\bigg(\sqrt{-2mE}\,\dsize{r'\over\hbar}\bigg)
  }{\alpha+\dsize{m\over\pi\hbar^2}\bigg[
  \ln\dsize{\sqrt{-2mE}\over2\hbar}-\Psi(1)\bigg]}
    \\   &\qquad
  +{m\over\pi\hbar^2}
  \sum_{l=1}^\infty\e^{\i l(\phi''-\phi')}
  K_l\bigg(\sqrt{-2mE}\,{r_>\over\hbar}\bigg)
  I_l\bigg(\sqrt{-2mE}\,{r_<\over\hbar}\bigg)
  \enspace.
  \tag\NUM.\num\endalign$$
\minus
\edef\numca{\NUM.\num}\plus%
and polar coordinates about $\vert\vec x-\vec a\vert$ have been used in
the last line. The one bound state wave-function
\plus$$\Psi^{(\alpha)}={1\over\sqrt{\pi}}
  K_0\big(\vert\vec x-\vec a\vert
                         \e^{-2[\alpha\pi\hbar^2/m-\Psi(1)]}\big)
  \tag\NUM.\num$$
has the energy
\plus$$
  E^{(\alpha)}=-{2\hbar^2\over m}\e^{-2[\alpha\pi\hbar^2/m-\Psi(1)]}
  \enspace.
  \tag\NUM.\num$$
The normalization of the bound state wave-function follows from the
integral representation [\GRA, p.672]
\plus$$\int_0^\infty xK_\nu(ax)K_\nu(bx)dx=
  \left\{\alignedat 3
  &\dsize{\pi(ab)^{-\nu}(a^{2\nu}-b^{2\nu})\over
   2\sin(\nu\pi)(a^2-b^2)}\enspace,
                   &\qquad  &\nu>0\enspace,  \\
  &\half\enspace,  &\qquad  &\nu=0\enspace.  \endalignedat\right.
  \tag\NUM.\num$$
\goodbreak

\subsubsection{The harmonic oscillator}
In order to discuss the harmonic oscillator we need the radial Green
function for the harmonic oscillator in $D$ dimensions. We have [\BVK,
\GOOb, \GRSb, \PI]
\plus$$\myalign
   &(r'r'')^{1-D\over2}\ih\int_0^\infty dT\,\e^{\i ET/\hbar}
         \\   &\qquad\times
   \int\limits_{r(t')=r'}^{r(t'')=r''}\CD r(t)
   \exp\left[\ih\int_{t'}^{t''}\bigg({m\over2}\dot r^2
  -\hbar^2{(l+{D-2\over2})^2-\viert\over2mr^2}
           -{m\over2}\omega^2r^2\bigg)dt\right]
         \\   &
   ={\Gamma[\half(l+{D\over2}
   -{E\over\hbar\omega})]\over\hbar\omega(r'r'')^{D/2}\Gamma(l+D/2)}
  W_{{E\over2\hbar\omega},\half(l+{D-2\over2})}
   \bigg({m\omega\over\hbar}r_>^2\bigg)
  M_{{E\over2\hbar\omega},\half(l+{D-2\over2})}
   \bigg({m\omega\over\hbar}r_<^2\bigg)\enspace.
         \\   &
  \tag\NUM.\num\endalign$$
We consider $D=2$. For the determination of the behaviour of the
Whittaker functions for $z\to0$ use has to been made of the
representations [\GRA, p.1059]
$$\myalign
  M_{\lambda,\mu}(z)&=z^{\mu+1/2}\e^{-z/2}
  {_1}F_1(\bhalf+\mu-\lambda;2\mu+1;z)\enspace,
  \tag\NUM.\num\\   \global\plus
  W_{\lambda,\mu}(z)&=
  {\Gamma(-2\mu)\over\Gamma(\half-\mu-\lambda)}M_{\lambda,\mu}(z)
  +{\Gamma(2\mu)\over\Gamma(\half+\mu-\lambda)}M_{\lambda,-\mu}(z)
  \tag\NUM.\num\endalign$$
\minus
\edef\numcg{\NUM.\num}\plus
respectively [\GRA, p.1063]
\plus$$\myalign
  W_{\lambda,\mu}(z)
  &={(-1)^{2\mu}z^{\mu+1/2}\e^{-z/2}\over
     \Gamma(\half-\mu-\lambda)\Gamma(\half+\mu-\lambda)}
  \Bigg\{\sum_{k=0}^\infty{\Gamma(\mu+k-\lambda+\half)\over k!(2\mu+k)!}
  z^k
  \\   &\qquad\qquad\qquad\times
     \Big[\Psi(k+1)+\Psi(2\mu+k+1)-\Psi(\mu+k-\lambda+\bhalf)-\ln z\Big]
  \\   &\qquad\qquad
  +(-z)^{-2\mu}\sum_{k=0}^{2\mu-1}
   {\Gamma(2\mu-k)\Gamma(k-\mu-\lambda+\half)\over k!}(-z)^k\Bigg\}
  \enspace.
  \tag\NUM.\num\endalign$$
\edef\numch{\NUM.\num}%
In the latter equation it is required that $2\mu\in\bbbn_0$, and
the last term is understood to be ignored for $\mu=0$. Putting the
interactions center at the coordinate origin, we set
\plus$$g(r)={\Gamma(\half-E/2\hbar\omega)\over2\pi\hbar\omega r}
   W_{E/2\hbar\omega,0}\bigg({m\omega\over\hbar}r^2\bigg)
   M_{E/2\hbar\omega,0}\bigg({m\omega\over\hbar}r^2\bigg)\enspace,
  \tag\NUM.\num$$
and therefore we obtain due to the general theory
\plus$$\myalign
  &\ih\int_0^\infty dT\,\e^{\i ET/\hbar}
  \int\limits_{\vec x(t')=\vec x'}^{\vec x(t'')=\vec x''}
  \CD_{\Gamma^{(\omega)}_{\alpha,\vec a}}^2\vec x(t)
  \exp\Bigg\{\ih\int_{t'}^{t''}\bigg[{m\over2}{\dot{\vec x}}^2
  -{m\over2}\omega^2{\vec x}^2
  +\gamma\delta(\vec x)\bigg]dt\Bigg\}
  \\   &
  ={\Gamma(\half-E/2\hbar\omega)\over2\pi\hbar\omega r'r''}
   W_{E/2\hbar\omega,0}\bigg({m\omega\over\hbar}r_>^2\bigg)
   M_{E/2\hbar\omega,0}\bigg({m\omega\over\hbar}r_<^2\bigg)
  \\   &\qquad
  +{m\Gamma^2(\half-E/2\hbar\omega)\over4\pi^2\hbar^3\omega r'r''}
   \dsize\thickfrac{
   W_{E/2\hbar\omega,0}\bigg(\dsize{m\omega\over\hbar}{r''}^2\bigg)
   W_{E/2\hbar\omega,0}\bigg(\dsize{m\omega\over\hbar}{r'}^2\bigg)
  }{\alpha+\dsize{m\over2\pi\hbar^2}\bigg[
  \Psi\bigg(\dsize\half-\dsize{E\over2\hbar\omega}\bigg)
   +\ln\dsize{m\omega\over\hbar}-2\Psi(1)\bigg]}
  \\   &\qquad
  +\sum_{l=1}^\infty\e^{\i l(\phi''-\phi')}
  {\Gamma[\half(l+1-E/\hbar\omega)]\over2\pi\hbar\omega l!\,r'r''}
   W_{E/2\hbar\omega,l/2}\bigg({m\omega\over\hbar}{r'}^2\bigg)
   M_{E/2\hbar\omega,l/2}\bigg({m\omega\over\hbar}{r''}^2\bigg)
  \enspace.
  \\   &
  \tag\NUM.\num\endalign$$
The bound state energy levels $E_n^{(\alpha)}$ are determined by
\plus$$\alpha+\dsize{m\over2\pi\hbar^2}\left[
  \Psi\Bigg(\dsize\half-\dsize{E_n^{(\alpha)}\over2\hbar\omega}\Bigg)
   +\ln\dsize{m\omega\over\hbar}-2\Psi(1)\right]=0\enspace.
  \tag\NUM.\num$$
\goodbreak

\subsubsection{The Coulomb potential}
In order to treat the two-dimensional Coulomb potential case, we must
know the radial two-dimensional Coulomb Green's function. We have
(c.f.\ the references in the one-dimensional case)
\plus$$\myalign
  & \ih \int_0^\infty  dT\,\e^{\i TE/\hbar}
  \int\limits_{\vec x(t')=\vec x'}^{\vec x(t'')=\vec x''}\CD^D\vec x(t)
  \exp\left[\ih\int_{t'}^{t''}\bigg({m\over2}
  {\dot{\vec x}}^2+{q_1q_2\over\vert\vec x\vert}\bigg)dt\right]
         \\   &
  =\overh\sum_{l=0}^\infty S_l^{(D)}(\Omega'')S_l^{(D)\,*}(\Omega')
  (r'r'')^{1-D\over2}\sqrt{-{m\over2E}}\,
   {\Gamma\big(l+{D-1\over2}-\kappa\big)\over(2l+D-2)! }
         \\   &\qquad\times
   W_{\kappa,l+{D-2\over2}}\bigg(\sqrt{-8mE}\,{r_>\over\hbar}\bigg)
   M_{\kappa,l+{D-2\over2}}\bigg(\sqrt{-8mE}\,{r_<\over\hbar}\bigg)
  \tag\NUM.\num\endalign$$
($\kappa=q_1q_2\sqrt{-m/2E}/\hbar$) and the $S_l^{(D)}(\Omega)$ are the
(real) hyperspherical harmonics on the $S^{(D-1)}$-sphere taken at the
unit vector $\Omega$ on the sphere. We consider $D=2$, and putting the
interactions center at the coordinate origin, we obtain in the
regularization procedure by means of (\numcg,\numch)
\plus$$g(r)={1\over2\pi\hbar}\sqrt{-{m\over2E}}
   {\Gamma(\half-\kappa)\over\sqrt{r}}
   W_{\kappa,0}\bigg(\sqrt{-8mE}\,{r\over\hbar}\bigg)
   M_{\kappa,0}\bigg(\sqrt{-8mE}\,{r\over\hbar}\bigg)\enspace.
  \tag\NUM.\num$$
Therefore we get for the $\delta$-perturbed two-dimensional Coulomb
potential problem
\plus$$\myalign
  &\ih\int_0^\infty dT\,\e^{\i ET/\hbar}
  \int\limits_{\vec x(t')=\vec x'}^{\vec x(t'')=\vec x''}
  \CD_{\Gamma^{(C)}_{\alpha,\vec a}}^2\vec x(t)
  \exp\Bigg\{\ih\int_{t'}^{t''}\bigg[{m\over2}{\dot{\vec x}}^2
  +{q_1q_2\over\vert\vec x\vert}
  +\gamma\delta(\vec x)\bigg]dt\Bigg\}
  \\   &
  =\sqrt{-{m\over2E}}{\Gamma(\half-\kappa)\over2\pi\hbar\sqrt{r'r''}}
   W_{\kappa,0}\bigg(\sqrt{-8mE}\,{r_>\over\hbar}\bigg)
   M_{\kappa,0}\bigg(\sqrt{-8mE}\,{r_<\over\hbar}\bigg)
  \\   &
  +\bigg({m\over\pi\hbar^2}\bigg)^2
   {\Gamma^2(\half-\kappa)\over\sqrt{r'r''}}
   \dsize\thickfrac{
   W_{\kappa,0}\bigg(\sqrt{-8mE}\,\dsize{r'\over\hbar}\bigg)
   W_{\kappa,0}\bigg(\sqrt{-8mE}\,\dsize{r''\over\hbar}\bigg)
  }{\alpha+\dsize{m\over2\pi\hbar^2}\bigg[
  \Psi\bigg(\dsize\half-\kappa\bigg)
   +\ln\dsize{\sqrt{-8mE}\over\hbar}-2\Psi(1)\bigg]}
  \\   &
  +{1\over2\pi\hbar}\sqrt{-{m\over2E}}
  \sum_{l=1}^\infty\e^{\i l(\phi''-\phi')}
   {\Gamma(l+\half-\kappa)\over(2l)!\,\sqrt{r'r''}}
   W_{\kappa,l}\bigg(\sqrt{-8mE}\,{r_>\over\hbar}\bigg)
   M_{\kappa,l}\bigg(\sqrt{-8mE}\,{r_<\over\hbar}\bigg)\enspace.
 \\   &
  \tag\NUM.\num\endalign$$
The bound state energy levels $E_n^{(\alpha)}$ are determined by
\plus$$\alpha+\dsize{m\over2\pi\hbar^2}\left[
  \Psi\Bigg(\dsize\half
    -{q_1q_2\over\hbar}\sqrt{-{m\over2E_n^{(\alpha)}}}\Bigg)
   +\ln\dsize{\sqrt{-8mE_n^{(\alpha)}}\over\hbar}-2\Psi(1)\right]=0
  \enspace.
  \tag\NUM.\num$$
\goodbreak

\subsubsection{Multiple $\delta$-function perturbations}
Similarly as in the one-dimensional case, we can consider the case of
multiple $\delta$-function perturbations in two dimensions, therefore
a lattice of $\delta$-function perturbations. Repeating the
regularization process for each additional $\delta$-function we obtain
by induction [\AGHHa, \AGHHb]
\plus$$\myalign
  &\ih\int_0^\infty dT\,\e^{\i ET/\hbar}
  \int\limits_{\vec x(t')=\vec x'}^{\vec x(t'')=\vec x''}
  \CD_{\Gamma^{(0)}_{\{\alpha\},\{\vec a\}}}^2\vec x(t)
  \exp\left\{\ih\int_{t'}^{t''}\left[{m\over2}{\dot{\vec x}}^2
  +\sum_{k=1}^N\gamma_k\delta(\vec x-\vec a_k)\right]dt\right\}
  \\   &
  ={m\over\pi\hbar^2}K_0\bigg({\sqrt{-2mE}\over\hbar}
         \vert\vec x''-\vec x'\vert\bigg)
   +\bigg({m\over\pi\hbar^2}\bigg)^2\sum_{j,j'=1}^N
  \Big(\Gamma^{(0)}_{\{\alpha\},\{\vec a\}}\Big)_{j,j'}^{-1}
  \\   &\qquad\qquad\qquad\times
  K_0\bigg({\sqrt{-2mE}\over\hbar}\vert\vec x''-\vec a_j\vert\bigg)
  K_0\bigg({\sqrt{-2mE}\over\hbar}\vert\vec a_{j'}-\vec x'\vert\bigg)
  \enspace.
  \tag\NUM.\num\endalign$$
with the matrix $\Gamma^{(0)}_{\{\alpha\},\{\vec a\}}$ given by $\big[\{
\alpha\}=\{\alpha_k\}_{k=1}^N$, $\{\vec a\}=\{\vec a_k\}_{k=1}^N$,
Note that $G_2^{(0)}(\vec x,\vec y;E)=G_2^{(0)}(\vec x-\vec y,0;E)\equiv
G_2^{(0)}(\vec x-\vec y;E)\big]$
$$\myalign
  \big(\Gamma^{(0)}_{\{\alpha\},\{\vec a\}}\big)_{j,j'}
       &
  =\bigg(\alpha_j-\Psi(1)+{m\over\pi\hbar^2}
   \ln\dsize{\sqrt{-2mE}\over2\hbar}\bigg)\delta_{jj'}
   -\widetilde G_2^{(0)}(\vec a_j,\vec a_{j'};E)
  \tag\NUM.\num\\   \global\plus
       &
  =\left\{\alignedat 3
  &\alpha_j+\lim_{\vert\vec x\vert\to0}
           \big[G_2^{(0)}(\vec x;0)-G_2^{(0)}(\vec x;E)\big]
  &\qquad  & j=j'  \\
  &-G_2^{(0)}(\vec a_j,\vec a_j';E)
  &\qquad  & j\not=j'  \\
  \endalignedat\right.
  \enspace,
  \tag\NUM.\num\endalign$$
where $G_2^{(0)}(\vec x;0)=-(m/\pi\hbar^2)\ln\vert\vec x\vert$ ($\vec x
\in\bbbr^2\setminus\{0\}$), and $\widetilde G_2^{(0)}(\vec x;E)=
G_2^{(0)}(\vec x;E)$ for $\vec x\not=0$, and $\widetilde G_2^{(0)}(\vec
x;E)=0$ otherwise. The bound states energy levels $E_n^{(\alpha)}$ are
determined by $\det\big(\Gamma^{(0)}_{\{\alpha\},\{\vec a\}}
(E_n^{(\alpha)})\big)=0$.
\goodbreak

\subsection{Three-dimensional examples}\nobreak\ssf
\subsubsection{The free particle}
Again we start with the free particle. Form (\numce) we know the free
particle Green function and we obtain according to the general theory
\plus$$g(r)={m\over2\pi\hbar^2}\e^{-r\sqrt{-2mE}/\hbar}\enspace,
  \tag\NUM.\num$$
($r=\vert\vec x\vert$, $\vec x\in\bbbr^3 \setminus\{0\}$).
Therefore we get (c.f.\ (\numbb) and [\AGHHb])
\plus$$\myalign
  &\ih\int_0^\infty dT\,\e^{\i ET/\hbar}
  \int\limits_{\vec x(t')=\vec x'}^{\vec x(t'')=\vec x''}
  \CD_{\Gamma^{(0)}_{\alpha,\vec a}}^3\vec x(t)
  \exp\Bigg\{\ih\int_{t'}^{t''}\bigg[{m\over2}{\dot{\vec x}}^2
  +\gamma\delta(\vec x-\vec a)\bigg]dt\Bigg\}
  \\   &
  =G_3^{(0)}(\vec x'',\vec x';E)+\dsize\thickfrac{
   G_3^{(0)}(\vec x'',\vec a;E)G_3^{(0)}(\vec a,\vec x';E)}
    {\alpha+\dsize{m\over2\pi\hbar^3}\sqrt{-2mE}}
  \enspace.
  \tag\NUM.\num\endalign$$
Expanding into three-dimensional polar coordinates we obtain
(compare also [\AGHHb])
\plus$$\myalign
  &{1\over r'r''}\ih\sum_{l=0}^\infty\int_0^\infty dT\,\e^{\i ET/\hbar}
  \int\limits_{r(t')=r'}^{r(t'')=r''}\mu_{l+\half}[r^2]
  \CD_{\Gamma^{(0)}_{\alpha,\vec a}} r(t)
  \exp\Bigg\{\ih\int_{t'}^{t''}\bigg[{m\over2}\dot r^2
                              +\gamma\delta(r)\bigg]dt\Bigg\}
  \\   &
  ={1\over2\pi\hbar r'r''}\sqrt{-{m\over2E}}
   \sinh\bigg(\sqrt{-2mE}\,{r_<\over\hbar}\bigg)
   \exp\bigg(-\sqrt{-2mE}\,{r_>\over\hbar}\bigg)
  \\   &\qquad
  +\bigg({m\over2\pi\hbar^2}\bigg)^2{1\over r'r''}
  \dsize\thickfrac{
   \exp\bigg(-\sqrt{-2mE}\,\dsize{r'+r''\over\hbar}\bigg)
   }{\alpha+\dsize{m\over2\pi\hbar^3}\sqrt{-2mE}}
  \\   &\qquad
  +\sum_{l=1}^\infty {m\over\hbar^2\sqrt{r'r''}}
   I_{l+\half}\bigg(\sqrt{-2mE}\,{r_<\over\hbar}\bigg)
   K_{l+\half}\bigg(\sqrt{-2mE}\,{r_>\over\hbar}\bigg)\enspace.
  \tag\NUM.\num\endalign$$

Denoting the free three-dimensional propagator by
\plus$$K^{(0)}(\vec x'',\vec x';T)
   =\bigg({m\over2\pi\i\hbar T}\bigg)^{3/2}
  \exp\bigg({\i m\over2\hbar T}\vert\vec x''-\vec x'\vert^2\bigg)
  \enspace,
  \tag\NUM.\num$$
it is possible to apply the inverse Laplace-Fourier transformation
and one obtains for the perturbed propagator [\SCATE]
\plus$$\myalign
  \int\limits_{\vec x(t')=\vec x'}^{\vec x(t'')=\vec x''}
  &\CD_{\Gamma^{(0)}_{\alpha,\vec a}}^3\vec x(t)
  \exp\Bigg\{\ih\int_{t'}^{t''}\bigg[{m\over2}{\dot{\vec x}}^2
  +\gamma\delta(\vec x-\vec a)\bigg]dt\Bigg\}
  \\   &
  =K^{(0)}_3(\vec x'',\vec x';T)
  +{1\over\vert\vec a-\vec x'\vert\vert\vec x''-\vec a\vert}
  \int_0^\infty \e^{-2\pi\alpha\hbar u/m}
  (u+\vert\vec a-\vec x'\vert+\vert\vec x''-\vec a\vert)
  \\   &\qquad\qquad\qquad\times
  K^{(0)}_3(u+\vert\vec a-\vec x'\vert+\vert\vec x''-\vec a\vert,0;T)du
  \enspace,
  \tag\NUM.\num a\\   &
  =K^{(0)}_3(\vec x'',\vec x';T)
   +{\i\hbar T\over m\vert\vec a-\vec x'\vert\vert\vec x''-\vec a\vert}
    K^{(0)}_3(\vert\vec a-\vec x'\vert+\vert\vec x''-\vec a\vert,0;T)
  \enspace,
  \tag\NUM.\num b\\   &
  =K^{(0)}_3(\vec x'',\vec x';T)
   +\Psi^{(\alpha)}(\vec x')\Psi^{(\alpha)}(\vec x'')
   \e^{\i E^{(\alpha)}T/\hbar}
  \\   &\qquad
  +{1\over\vert\vec a-\vec x'\vert\vert\vec x''-\vec a\vert}
    \int_0^\infty \e^{-2\pi\vert\alpha\vert\hbar u/m}
    (u-\vert\vec a-\vec x'\vert-\vert\vec x''-\vec a\vert)
  \\   &\qquad\qquad\qquad\times
   K^{(0)}_3(u-\vert\vec a-\vec x'\vert-\vert\vec x''-\vec a\vert,0;T)du
  \enspace,
  \tag\NUM.\num c\endalign$$
for $\alpha>0$, $\alpha=0$ and $\alpha<0$, respectively. For
$\alpha<0$ there is one bound state wave-function
\plus$$\Psi^{(\alpha)}(\vec x)=\sqrt{-{\alpha\hbar^2\over m}}\,
   {\e^{-2\pi\alpha\hbar\vert\vec x-\vec a\vert/m}
    \over\vert\vec x-\vec a\vert}
  \enspace,
  \tag\NUM.\num$$
and the energy eigen-value $E^{(\alpha)}$ has the form
\plus$$E^{(\alpha)}=-2\pi^2{\alpha^2\hbar^6\over m^3}\enspace.
  \tag\NUM.\num$$
\goodbreak

\subsubsection{The harmonic oscillator}
In order to discuss the harmonic oscillator we use the representation
(\numcd). Applying the regularization procedure and by means of
(\numcg,\numch) we set
\plus$$g(r)={\Gamma[\half(3/2-E/\hbar\omega)]\over4\pi\hbar\omega r^2}
   W_{E/2\hbar\omega,1/4}\bigg({m\omega\over\hbar}r^2\bigg)
   M_{E/2\hbar\omega,1/4}\bigg({m\omega\over\hbar}r^2\bigg)
   \enspace.
  \tag\NUM.\num$$
We then obtain by denoting the three-dimensional harmonic oscillator
Green function  by $G^{(\omega)}(E)$ from (\numcd) ($\nu=-\half+E/\hbar
\omega$, $a=\vert\vec a\vert$, $\tilde a=\sqrt{2m\omega/\hbar}\,a$,
compare also [\JAMA]).
$$\myalign
  &\ih\int_0^\infty dT\,\e^{\i ET/\hbar}
  \int\limits_{\vec x(t')=\vec x'}^{\vec x(t'')=\vec x''}
  \CD_{\Gamma^{(\omega)}_{\alpha,\vec a}}^3\vec x(t)
  \exp\Bigg\{\ih\int_{t'}^{t''}\bigg[{m\over2}{\dot{\vec x}}^2
  -{m\over2}\omega^2{\vec x}^2
  +\gamma\delta(\vec x-\vec a)\bigg]dt\Bigg\}
  \\   &
  =G^{(\omega)}(\vec x'',\vec x';E)
  +{G^{(\omega)}(\vec x'',\vec a;E)G^{(\omega)}(\vec a,\vec x';E)
    \over\Gamma^{(\omega)}_{\alpha,\vec a}(E)}\enspace,
  \tag\NUM.\num a\\
       &
  \text{where}
  \\   &\quad
  \Gamma^{(\omega)}_{\alpha,\vec a}=
  \alpha+{m\Gamma(-\nu)\over2(2\pi)^{3/2}\hbar^2}
  \Bigg\{{1\over a}\Big[D_\nu(\tilde a)D'_\nu(-\tilde a)
                        -D'_\nu(\tilde a)D_\nu(-\tilde a)\Big]
  \\   &\qquad\qquad
  +\sqrt{2m\omega\over\hbar}\Big[
  D''_\nu(\tilde a)D_\nu(-\tilde a)
  +2D'_\nu(\tilde a)D'_\nu(-\tilde a)
  +D_\nu(\tilde a)D''_\nu(-\tilde a)\Big]\Bigg\}\enspace.\qquad
  \tag\NUM.\num b\\ \global\plus
       &
  \text{(The case $\vec a=\vec 0$ gives:)}
  \\   &
  ={\Gamma[\half(3/2-E/\hbar\omega)]\over4\pi\hbar\omega(r'r'')^{3/2}}
   W_{E/2\hbar\omega,1/4}\bigg({m\omega\over\hbar}r_>^2\bigg)
   M_{E/2\hbar\omega,1/4}\bigg({m\omega\over\hbar}r_<^2\bigg)
  \\   &\qquad
  +{m^2\Gamma^2[\half(3/2-E/\hbar\omega)]2^{-(E/\hbar\omega-\half)}
   \over4\pi^3\hbar^4r'r''}
  \\   &\qquad\qquad\qquad\times
   \dsize\thickfrac{
   D_{E/\hbar\omega-\half}\bigg(\dsize\sqrt{2m\omega\over\hbar}r''\bigg)
   D_{E/\hbar\omega-\half}\bigg(\dsize\sqrt{2m\omega\over\hbar}r'\bigg)
   }{\alpha-\dsize{m\over2\pi\hbar^2}\dsize\sqrt{m\omega\over\hbar}\,
   \dsize{\Gamma[\half(3/2-E/\hbar\omega)]
           \over\Gamma[\half(1/2-E/\hbar\omega)]}}
  \\   &\qquad
  +\sum_{l=1}^\infty{\Gamma[\half(l+3/2-E/\hbar\omega)]\over
    \hbar\omega\Gamma(l+3/2)(r'r'')^2}
  \sum_{n=-l}^lY_l^{n\,*}(\theta',\phi')Y_l^n(\theta'',\phi'')
  \\   &\qquad\qquad\qquad\times
  W_{E/2\hbar\omega,\half(l+\half)}\bigg({m\omega\over\hbar}{r'}^2\bigg)
 M_{E/2\hbar\omega,\half(l+\half)}\bigg({m\omega\over\hbar}{r''}^2\bigg)
  \enspace.
  \tag\NUM.\num\endalign$$
For $\vec a=\vec 0$ use has been made of $D_\nu(0)=\sqrt{\pi}\,2^{\nu/2}
/\Gamma({1-\nu\over2})$, and $D'_\nu(z)=\nu D_{\nu-1}(z)-zD_\nu(z)/2$,
and $Y_l^n(\theta,\phi)$ are the spherical harmonics on the sphere.
 The bound states energy levels $E_n^{(\alpha)}$ are determined by
$\Gamma^{(\omega)}_{\alpha,\{\vec a\}}(E_n^{(\alpha)})=0$.
\goodbreak

\subsubsection{The Coulomb potential}
Let us denote by $G^{(C)}(E)$ the three-dimensional Coulomb Green
function as given by Hostler [\HOST] ($\kappa=q_1q_2\sqrt{-m/2E}/\hbar$)
\plus$$\myalign
    \ih&\int_0^\infty  dT\,\e^{\i TE/\hbar}
  \int\limits_{\vec x(t')=\vec x'}^{\vec x(t'')=\vec x''}\CD^3\vec x(t)
  \exp\left[\ih\int_{t'}^{t''}
  \bigg({m\over2}{\dot{\vec x}}^2
   +{q_1q_2\over\vert\vec x\vert}\bigg)dt\right]
         \\   &
  =-{m\Gamma(1-\kappa)\over2\pi\hbar^2\vert\vec x''-\vec x'\vert}
   \left\vert\matrix
   W_{\kappa,1/2}\bigg(\sqrt{-8mE}\,\dsize{x_>\over\hbar}\bigg)
  &M_{\kappa,1/2}\bigg(\sqrt{-8mE}\,\dsize{x_<\over\hbar}\bigg) \\
   W_{\kappa,1/2}'\bigg(\sqrt{-8mE}\,\dsize{x_>\over\hbar}\bigg)
  &M_{\kappa,1/2}'\bigg(\sqrt{-8mE}\,\dsize{x_<\over\hbar}\bigg)
  \endmatrix\right\vert\qquad
         \\   &
  =\sum_{l=0}^\infty\sum_{n=-l}^l
   Y^{n\,*}_l(\theta',\phi')Y^{n}_l(\theta'',\phi'')
   {1\over r'r''}\overh\sqrt{-{m\over2E}}
   {\Gamma(1+l-\kappa)\over(2l+1)!}
         \\   &\qquad\times
   W_{\kappa,l+\half}\left(\sqrt{-8mE}\,{r_>\over\hbar}\right)
   M_{\kappa,l+\half}\left(\sqrt{-8mE}\,{r_<\over\hbar}\right)
  \enspace.
  \tag\NUM.\num\endalign$$
According to the general theory we set (c.f.\ (\numcg,\numch) and
[\AGHHb])
\plus$$g(r)=\sqrt{-{m\over2E}}{\Gamma(1-\kappa)\over4\pi\hbar r}
   W_{\kappa,\half}\bigg(\sqrt{-8mE}\,{r\over\hbar}\bigg)
   M_{\kappa,\half}\bigg(\sqrt{-8mE}\,{r\over\hbar}\bigg)\enspace,
  \tag\NUM.\num$$
Therefore we obtain (compare also [\AGHHb, \KOM])
$$\myalign
  &\ih\int_0^\infty dT\,\e^{\i ET/\hbar}
  \int\limits_{\vec x(t')=\vec x'}^{\vec x(t'')=\vec x''}
  \CD_{\Gamma^{(C)}_{\alpha,\vec a}}^3\vec x(t)
  \exp\Bigg\{\ih\int_{t'}^{t''}\bigg[{m\over2}{\dot{\vec x}}^2
  +{q_1q_2\over\vert\vec x\vert}
  +\gamma\delta(\vec x-\vec a)\bigg]dt\Bigg\}
  \\   &
  =G^{(C)}(\vec x'',\vec x';E)
  +{G^{(C)}(\vec x'',\vec a;E)G^{(C)}(\vec a,\vec x';E)
    \over\Gamma^{(C)}_{\alpha,\vec a}(E)}\enspace,
  \tag\NUM.\num a\\
       &
  \text{where}
  \\   &\quad
   \Gamma^{(C)}_{\alpha,\vec a}(E)=
  \alpha+{m\Gamma(1-\kappa)\over2\pi\hbar^3}\sqrt{-8mE}
  \\   &\qquad\times
   \Big[2W'_{\kappa,\half}(2\tilde a)M'_{\kappa,\half}(2\tilde a)
   -W_{\kappa,\half}(2\tilde a)M''_{\kappa,\half}(2\tilde a)
   -M_{\kappa,\half}(2\tilde a)W''_{\kappa,\half}(2\tilde a)\Big]
  \enspace.
  \tag\NUM.\num b\\ \global\plus
       &
  \text{(The case $\vec a=\vec 0$ gives:)}
  \\   &
  =\sqrt{-{m\over2E}}{\Gamma(1-\kappa)\over4\pi\hbar r'r''}
   W_{\kappa,\half}\bigg(\sqrt{-8mE}\,{r_>\over\hbar}\bigg)
   M_{\kappa,\half}\bigg(\sqrt{-8mE}\,{r_<\over\hbar}\bigg)
  \\   &\qquad
  +\dsize\thickfrac{
   \bigg(\dsize{m\over2\pi\hbar^2}\bigg)^2
   \dsize{\Gamma^2(1-\kappa)\over r'r''}
   W_{\kappa,\half}\bigg(\sqrt{-8mE}\,\dsize{r'\over\hbar}\bigg)
   W_{\kappa,\half}\bigg(\sqrt{-8mE}\,\dsize{r''\over\hbar}\bigg)
   }{\alpha+\dsize{mq_1q_2\over2\pi\hbar^2}\bigg[
   {2m\over\hbar^2}\bigg(\Psi(1)+\Psi(2)-\Psi(1-\kappa)
    +\ln{\hbar q_1q_2\over\sqrt{-8mE}}\bigg)
    -{\sqrt{-2mE}\over\hbar q_1q_2}\,\bigg]}
  \\   &\qquad
  +{1\over\hbar}\sqrt{-{m\over2E}}
  \sum_{l=1}^\infty{\Gamma(l+1-\kappa)\over(2l+1)!\,r'r''}
  \sum_{n=-l}^lY_l^{n\,*}(\theta',\phi')Y_l^n(\theta'',\phi'')
  \\   &\qquad\qquad\qquad\times
   W_{\kappa,l+1/2}\bigg(\sqrt{-8mE}\,{r_>\over\hbar}\bigg)
   M_{\kappa,l+1/2}\bigg(\sqrt{-8mE}\,{r_<\over\hbar}\bigg)
  \enspace,
  \tag\NUM.\num\endalign$$
where three-dimensional polar coordinates have been used in the last
line. The bound state energy levels $E_n^{(\alpha)}$ are determined by
$\Gamma^{(C)}_{\alpha,\{\vec a\}}(E_n^{(\alpha)})=0$.
\goodbreak

\subsubsection{Multiple $\delta$-function perturbations}
As the last example in this sequel we again consider the case of
multiple $\delta$-function perturbations. Similarly as before we get by
induction [\AGHHb]
\plus$$\myalign
  &\ih\int_0^\infty dT\,\e^{\i ET/\hbar}
  \int\limits_{\vec x(t')=\vec x'}^{\vec x(t'')=\vec x''}
  \CD_{\Gamma^{(0)}_{\{\alpha\},\{\vec a\}}}^3\vec x(t)
  \exp\left\{\ih\int_{t'}^{t''}\left[{m\over2}{\dot{\vec x}}^2
  +\sum_{k=1}^N\gamma_k\delta(\vec x-\vec a_k)\right]dt\right\}
  \\   &
  =G_3^{(0)}(\vec x'',\vec x';E)+\sum_{j,j'=1}^N
  \Big(\Gamma^{(0)}_{\{\alpha\},\{\vec a\}}\Big)_{j,j'}^{-1}
  G_3^{(0)}(\vec x'',\vec a_j;E)G_3^{(0)}(\vec a_{j'},\vec x';E)
  \enspace.
  \tag\NUM.\num\endalign$$
with the matrix $\Gamma^{(0)}_{\{\alpha\},\{\vec a\}}$ given by $\big[\{
\alpha\}=\{\alpha_k\}_{k=1}^N$, $\{\vec a\}=\{\vec a_k\}_{k=1}^N$,
note that $G_3^{(0)}(\vec x,\vec y;E)=G_3^{(0)}(\vec x-\vec y,0;E)\equiv
G_3^{(0)}(\vec x-\vec y;E)\big]$
$$\myalign
  \big(\Gamma^{(0)}_{\{\alpha\},\{\vec a\}}\big)_{j,j'}
       &
  =\bigg(\alpha_j+{m\over2\pi\hbar^3}\sqrt{-2mE}\bigg)\delta_{jj'}
   -\widetilde G^{(0)}_3(\vec a_j,\vec a_{j'};E)
  \tag\NUM.\num\\   \global\plus
       &
  =\left\{\alignedat 3
  &\alpha_j
   +\lim_{\vert\vec x\vert\to0}
    \big[G_3^{(0)}(\vec x;0)-G_3^{(0)}(\vec x;E)\big]
  &\qquad  & j=j'    \enspace,  \\
  &-G_3^{(0)}(\vec a_j,\vec a_j';E)
  &\qquad  & j\not=j'\enspace,  \\
  \endalignedat\right.\qquad
  \tag\NUM.\num\endalign$$
where $G_3^{(0)}(\vec x;0)=m/2\pi\hbar^2\vert\vec x\vert$ ($\vec x\in
\bbbr^3\setminus\{0\}$), and $\widetilde G_3^{(0)}(\vec x;E)=G_3^{(0)}
(\vec x;E)$ for $\vec x\not=0$, and $\widetilde G_3^{(0)}(\vec x;E)=0$
otherwise. The bound states energy levels $E_n^{(\alpha)}$ are
determined by $\det\big(\Gamma^{(0)}_{\{\alpha\},\{\vec a\}}
(E_n^{(\alpha)})\big)=0$.
\goodbreak

\subsection{Nonstandard three-dimensional potentials}
In this subsection we want to discuss several potential problems
important in nuclear physics. They are
\medskip
\item{i)} the potential well,
\item{ii)} the Wood-Saxon potential,
\item{iii)} and the rotating Morse oscillator.
\medskip

These potentials can serve as models for the potential strength
in the vicinity of the nucleus more or less approximating the
strong interaction force. It is assumed that they are radial symmetric,
which allows an explicit solution for $s$-waves.
\goodbreak

\subsubsection{The potential step}
We first discuss the potential step. The Green function for the
radial potential step potential $V(r)=\Theta(b-r)V_0$ can be deduced
from the smooth-step potential (see below) and has the form [\GROx]
($k^2=2m(E+V_0)/\hbar^2$, $\chi^2=-2mE/\hbar^2$, $l=0$)
\plus$$\myalign
  & {\i\over4\pi \hbar r'r''}
   \int_0^\infty  dT\,\e^{\i TE/\hbar}
  \int\limits_{r(t')=x'}^{r(t'')=r''}\CD r(t)
  \exp\left[\ih\int_{t'}^{t''}\bigg({m\over2}\dot r^2
                         -[\Theta(r-b)-1]V_0\bigg)dt\right]
         \\   &
   ={\Theta(b-r')\Theta(b-r'')\over4\pi\hbar r'r''}
   \sqrt{-{m\over2(E+V_0)}}\,
         \\   &\qquad\times
   \e^{-\i k(r_<-b)}\bigg(\e^{\i k(r_>-b)}
    -{\chi+\i k\over\chi-\i k}\e^{-\i k(r_>-b)}\bigg)
         \\   &
   +{\Theta(r'-b)\Theta(r''-b)\over4\pi\hbar r'r''}
   \sqrt{-{m\over2E}}\,
   \e^{-\chi(r_>-b)}\bigg(\e^{\chi(r_<-b)}
    +{\chi+\i k\over\chi-\i k}\e^{-\chi(r_<-b)}\bigg)
         \\   &
   +{\Theta(r_>-b)\Theta(b-r_<)\over4\pi\hbar r'r''}
   {\sqrt{2m}\over\sqrt{-E}\,+\sqrt{-E-V_0}}
   \e^{-\i k(r_<-b)} \e^{-\chi(r_>-b)}
   \enspace,
  \tag\NUM.\num\endalign$$
[alternatively we can write $\e^{2\i\arctan(k/\chi)}=(\chi+\i k)/
\chi-\i k)\equiv\rho$]. Let us denote this Green function by
$G^{(PS)}(r'',r';E)$. According to the theory we set
\plus$$g(r)={1\over4\pi r\hbar}\sqrt{-{m\over2(E+V_0)}}\e^{-\i k(r-b)}
   \Big(\e^{\i k(r-b)}-\rho\e^{-\i k(r-b)}\Big)
   \bigg(1-{\e^{2\i k(r-b)}-\rho\over\e^{-2\i kb}-\rho}\bigg)\enspace.
  \tag\NUM.\num$$
Therefore we obtain
\hfuzz=6pt
\plus$$\myalign
  & {1\over4\pi r'r''}
    \ih \int_0^\infty  dT\,\e^{\i TE/\hbar}
  \int\limits_{r(t')=x'}^{r(t'')=r''}\CD_{\Gamma^{(PS)}_\alpha} r(t)
  \exp\left[\ih\int_{t'}^{t''}\bigg({m\over2}\dot r^2
                         -[\Theta(r-b)-1]V_0\bigg)dt\right]
         \\   &
  =G^{(PS)}(r'',r';E)+\big(\Gamma^{(PS)}_{\alpha,0}(E)\big)^{-1}
  G^{(PS)}(r'',0;E)G^{(PS)}(0,r';E)\enspace,
  \tag\NUM.\num\endalign$$
\hfuzz=3pt
where
\plus$$\Gamma^{(PS)}_\alpha(E)=\alpha+{m\over\pi\hbar^3}
  \dsize\thickfrac{
  \sqrt{-2m(E+V_0)}\,\dsize\frac{\sqrt{-E}\,-\sqrt{-E-V_0}}
                   {\sqrt{-E}\,+\sqrt{-E-V_0}}  }{
  \e^{2b\sqrt{-2m(E+V_0)}/\hbar}-\dsize\frac{\sqrt{-E}\,-\sqrt{-E-V_0}}
                   {\sqrt{-E}\,+\sqrt{-E-V_0}}
  }\enspace,
  \tag\NUM.\num$$
and $\Gamma^{(PS)}_\alpha(E_n^{(\alpha)})=0$ determines the bound state
energy levels.
\goodbreak

\subsubsection{The Wood-Saxon potential}
The Wood-Saxon potential can be seen as a smooth step-potential modeling
the potential strength in the vicinity of a nucleus [\FLU]. It is a
special case of the so-called modified P\"osch-Teller potentials [\PT].
The path integral solution of the modified P\"oschl-Teller potential
can be done via the path integral solution of the $\SU(1,1)$-path
integral [\BJ], form which the path integral solution of the
smooth-step potential can be derived [\GROe].

{}From the Green function representation of the modified P\"oschl-Teller
potential, respectively the Wood-Saxon potential [\GROs, \KLEMUS],
\plus$$\myalign
   \ih&                   \int_0^\infty dT\,\e^{\i ET/\hbar}
  \int\limits_{r(t')=r'}^{r(t'')=r''}\CD r(t)
  \exp\left\{\ih\int_{t'}^{t''}\bigg[{m\over2}\dot r^2
      +{V_0\over 1+\e^{(r-b)/R}}\bigg]dt\right\}
  \\   &
  ={2mR\over\hbar^2}{\Gamma(m_1)\Gamma(m_1+1)\over
           \Gamma(m_1+m_2+1)\Gamma(m_1-m_2+1)}
  \\   &\qquad\times
  \bigg({1-\tanh{r_<-b\over2R}\over2}\bigg)^{m_1-m_2\over2}
  \bigg({1+\tanh{r_<-b\over2R}\over2}\bigg)^{m_1+m_2\over2}
  \\   &\qquad\times
  \bigg({1-\tanh{r_>-b\over2R}\over2}\bigg)^{m_1-m_2\over2}
  \bigg({1+\tanh{r_>-b\over2R}\over2}\bigg)^{m_1+m_2\over2}
  \\   &\qquad\times
  {_2}F_1\bigg(m_1,m_1+1;m_1-m_2+1;{1-\tanh{r_>-b\over2R}\over2}\bigg)
  \\   &\qquad\times
  {_2}F_1\bigg(m_1,m_1+1;m_1+m_2+1;{1+\tanh{r_<-b\over2R}\over2}\bigg)
  \tag\NUM.\num\endalign$$
($m_{1,2}=\sqrt{2m}\,R\big(\sqrt{-E-V_0}\pm\sqrt{-E}\,\big)/\hbar$) we
then can state the Green function of the rotating radial Wood-Saxon
oscillator which has for $s$-waves the following form [\GROw]
\plus$$\myalign
   &{\i\over4\pi\hbar r'r''}
  \int_0^\infty dT\,\e^{\i ET/\hbar}
  \int\limits_{r(t')=r'}^{r(t'')=r''}\CD r(t)
  \exp\left[\ih\int_{t'}^{t''}\bigg({m\over2}\dot r^2
      +{V_0\over 1+\e^{(r-b)/R}}\bigg)dt\right]
         \\   &
  ={mR\over2\pi\hbar^2r'r''}{\Gamma(m_1)\Gamma(m_1+1)\over
           \Gamma(m_1+m_2+1)\Gamma(m_1-m_2+1)}
         \\   &\qquad\times
  \bigg({1-\tanh{r_<-b\over2R}\over2}\bigg)^{m_1-m_2\over2}
  \bigg({1+\tanh{r_<-b\over2R}\over2}\bigg)^{m_1+m_2\over2}
         \\   &\qquad\times
  \bigg({1-\tanh{r_>-b\over2R}\over2}\bigg)^{m_1-m_2\over2}
  \bigg({1+\tanh{r_>-b\over2R}\over2}\bigg)^{m_1+m_2\over2}
         \\   &\qquad\times\left\{
  {_2}F_1\bigg(m_1,m_1+1;m_1-m_2+1;{1-\tanh{r_>-b\over2R}\over2}\bigg)
  \right.\\   &\qquad\qquad\times
  {_2}F_1\bigg(m_1,m_1+1;m_1+m_2+1;{1+\tanh{r_<-b\over2R}\over2}\bigg)
         \\   &\qquad\qquad\quad-
  \dsize\thickfrac{{_2}F_1\bigg(m_1,m_1+1;m_1+m_2+1;
                     \dsize{1-\tanh{b\over2R}\over2}\bigg)}
  {{_2}F_1\bigg(m_1,m_1+1;m_1-m_2+1;
                             \dsize{1+\tanh{b\over2R}\over2}\bigg)}
         \\   &\qquad\qquad\qquad\times
  {_2}F_1\bigg(m_1,m_1+1;m_1-m_2+1;{1-\tanh{r'-b\over2R}\over2}\bigg)
         \\   &\qquad\qquad\qquad\times\left.
  {_2}F_1\bigg(m_1,m_1+1;m_1-m_2+1;{1-\tanh{r''-b\over2R}\over2}\bigg)
  \right\}\enspace,
  \tag\NUM.\num\endalign$$
Let us denote this Green function by $G^{(WS)}_0(r'',r';E)$. Following
the general theory we set $g(r)=rG^{(WS)}_0(r,r;E)$. Let us abbreviate
\plus$$f_{\pm}(r)=
  {_2}F_1\bigg(m_1,m_1+1;m_1\pm m_2+1;
      {1\pm\tanh{r''-b\over2R}\over2}\bigg)\enspace.
  \tag\NUM.\num$$
Note the relations
\minus$$\myalign
  f_{\pm}(0)&=
  {_2}F_1\bigg(m_1,m_1+1;m_1\pm m_2+1;
      {1\mp\tanh{b\over2R}\over2}\bigg)\enspace,
  \tag\NUM.\num\\   \global\plus
  f_{\pm}'(0)&=
  \pm{{_2}F_1'\big(m_1,m_1+1;m_1\pm m_2+1;\half(
      1\mp\tanh{b\over2R})\big)\over4R\cosh^2{b\over2R}}
  \enspace,
  \tag\NUM.\num\\   \global\plus
  f_{\pm}''(0)&=
  {{_2}F_1''\big(m_1,m_1+1;m_1\pm m_2+1;\half(
      1\mp\tanh{b\over2R})\big)\over16R^2\cosh^4{b\over2R}}
  \\   &\qquad
  \mp\tanh{b\over2R}
     {{_2}F_1'\big(m_1,m_1+1;m_1\pm m_2+1;\half(
      1\mp\tanh{b\over2R})\big)\over4R^2\cosh^2{b\over2R}}
  \enspace.
  \tag\NUM.\num\endalign$$
We now find
$$\myalign
  g_{0,3}^{(WS)}&={R\,\Gamma(m_1)\Gamma(m_1+1)\over
           \Gamma(m_1+m_2+1)\Gamma(m_1-m_2+1)}
  \\   &\qquad\times
  \bigg({1+\tanh{b\over2R}\over2}\bigg)^{m_1-m_2}
  \bigg({1-\tanh{b\over2R}\over2}\bigg)^{m_1+m_2}
  \Big[f_-(0)f'_+(0)-f_-'(0)f_+(0)\Big]
  \tag\NUM.\num\\   \global\plus
  g_{1,3}^{(WS)}&={R\,\Gamma(m_1)\Gamma(m_1+1)\over
           \Gamma(m_1+m_2+1)\Gamma(m_1-m_2+1)}
  \\   &\qquad\times
  \bigg({1+\tanh{b\over2R}\over2}\bigg)^{m_1-m_2}
  \bigg({1-\tanh{b\over2R}\over2}\bigg)^{m_1+m_2}
         \\   &\qquad\times
  \bigg[\half\Big(f_-(0)f''_+(0)-f_-''(0)f_+(0)\Big)
  +f_-'(0)f_+'(0)-{f_+(0)\over f_-(0)}{f_-'}^2(0)\bigg]\enspace.
  \tag\NUM.\num\endalign$$
Therefore we obtain for the $\delta$-perturbed Wood-Saxon potential
\plus$$\myalign
   {\i\over4\pi\hbar r'r''}
             &
  \int_0^\infty dT\,\e^{\i ET/\hbar}
  \int\limits_{r(t')=r'}^{r(t'')=r''}\CD_{\Gamma_{\alpha,0}^{(WS)}} r(t)
  \exp\left[\ih\int_{t'}^{t''}\bigg({m\over2}\dot r^2
      +{V_0\over 1+\e^{(r-b)/R}}\bigg)dt\right]
         \\   &
  =G^{(WS)}(r'',r';E)+\big(\Gamma_{\alpha,0}^{(WS)}(E)\big)^{-1}
  G^{(WS)}(r'',0;E)G^{(WS)}(0,r';E)  \enspace,
  \tag\NUM.\num\endalign$$
with $\Gamma_{\alpha,0}^{(WS)}(E)$ given by
\plus$$\Gamma_{\alpha,0}^{(WS)}(E)=\alpha g_{0,3}^{(WS)}-g_{1,3}^{(WS)}
  \enspace,
  \tag\NUM.\num$$
and the bound-state energy levels $E_n$ are determined by
$\Gamma_{\alpha,0}^{(WS)}(E_n^{(\alpha)})=0$.
\goodbreak

\subsubsection{The rotating Morse oscillator}
Another model for a nuclear potential is the so-called rotating Morse
oscillator [\RUND]. Its Green function can be derived by path
integration via means of the Green function of the Morse potential
[\CIWI, \DURa, \GROb, \PAKSa] and has for $s$-waves the following form
[\GROw]
\plus$$\myalign
  &{\i\over4\pi\hbar r'r''}
  \int_0^\infty dT\,\e^{\i ET/\hbar}
  \int\limits_{r(t')=r'}^{r(t'')=r''}\CD r(t)
  \exp\left\{\ih\int_{t'}^{t''}\bigg[{m\over2}\dot r^2
  -{V_0^2\hbar^2\over2m}(\e^{-2r}-2\alpha\e^{-r})\bigg]dt\right\}
  \\   &
  ={m\over4\pi V_0\hbar^2r'r''}
  {\Gamma(\half+\sqrt{-2mE}/\hbar-\alpha V_0)\over
   \Gamma(1+\sqrt{-8mE}/\hbar)\e^{(r'+r'')/2}}
         \\   &\quad\times
  \left\{
  W_{\alpha V_0,\sqrt{-2mE}/\hbar}\big(2V_0 \e^{-r_<}\big)
  M_{\alpha V_0,\sqrt{-2mE}/\hbar}\big(2V_0 \e^{-r_>}\big)
\vphantom{\dfrac{W_{\alpha V_0,\sqrt{-2mE}/\hbar}(2V_0)}
  {M_{\alpha V_0,\sqrt{-2mE}/\hbar}(2V_0)}}
  \right.\\   &\left.\qquad\quad-
  \dsize\thickfrac{M_{\alpha V_0,\sqrt{-2mE}/\hbar}(2V_0)}
  {W_{\alpha V_0,\sqrt{-2mE}/\hbar}(2V_0)}
  M_{\alpha V_0,\sqrt{-2mE}/\hbar}\big(2V_0 \e^{-r'}\big)
  M_{\alpha V_0,\sqrt{-2mE}/\hbar}\big(2V_0 \e^{-r''}\big)
  \right\}\enspace.
         \\   &
  \tag\NUM.\num\endalign$$
Let us denote this Green function by $G^{(M)}_0(r'',r';E)$.
\plus$$ w_1(r)=M_{\alpha V_0,\sqrt{-2mE}/\hbar}(2V_0\e^{-r})
   \enspace,\qquad
   w_2(r)=W_{\alpha V_0,\sqrt{-2mE}/\hbar}(2V_0\e^{-r})
   \enspace.
  \tag\NUM.\num$$
Note the relations
$$\myalign
  w_1(0)&=
  M_{\alpha V_0,\sqrt{-2mE}/\hbar}(2V_0)\enspace,\qquad
   w_1'(0)=
  -2V_0M_{\alpha V_0,\sqrt{-2mE}/\hbar}'(2V_0)\enspace,
  \tag\NUM.\num\\   \global\plus
   w_1''(0)&=
   2V_0M_{\alpha V_0,\sqrt{-2mE}/\hbar}'(2V_0)
  +4V_0^2M_{\alpha V_0,\sqrt{-2mE}/\hbar}''(2V_0)\enspace.
  \tag\NUM.\num\endalign$$
According to the theory we set $g(r)=rG^{(M)}(r,r;E)$. Therefore we
obtain
$$\myalign
  g_{0,3}^{(M)}&=
  {\Gamma(\half+\sqrt{-2mE}/\hbar-\alpha V_0)\over
   2V_0\Gamma(1+\sqrt{-8mE}/\hbar)}
  \Big[w_1(0)w_2'(0)-w_1'(0)w_2(0)\Big]
  \tag\NUM.\num\\   \global\plus
  g_{1,3}^{(M)}&=
  {\Gamma(\half+\sqrt{-2mE}/\hbar-\alpha V_0)\over
   \Gamma(1+\sqrt{-8mE}/\hbar)}
  \\   &\qquad\times
  \bigg[\half\Big(w_1(0)w_2''(0)-w_1''(0)w_2(0)\Big)
    +w_1'(0)w_2'(0)-{w_2(0)\over w_1(0)}{w_1'}^2(0)\bigg]\enspace.
  \tag\NUM.\num\endalign$$
\eject\noindent
Therefore we obtain for the $\delta$-perturbed rotating Morse oscillator
\plus$$\myalign
  &{\i\over4\pi\hbar r'r''}
  \int_0^\infty dT\,\e^{\i ET/\hbar}
  \\   &\qquad\times
  \int\limits_{r(t')=r'}^{r(t'')=r''}\CD_{\Gamma_{\alpha,0}^{(M)}} r(t)
  \exp\left\{\ih\int_{t'}^{t''}\bigg[{m\over2}\dot r^2
  -{V_0^2\hbar^2\over2m}(\e^{-2r}-2\alpha\e^{-r})\bigg]dt\right\}
  \\   &
  =G^{(M)}(r'',r';E)+\big(\Gamma_{\alpha,0}^{(M)}(E)\big)^{-1}
  G^{(M)}(r'',0;E)G^{(M)}(0,r';E) \enspace,
  \tag\NUM.\num\endalign$$
with $\Gamma_{\alpha,0}^{(M)}(E)$ given by
\plus$$\Gamma_{\alpha,0}^{(M)}(E)=\alpha
  g_{0,3}^{(M)}-g_{1,3}^{(M)}\enspace,
  \tag\NUM.\num$$
and the bound-state energy levels $E_n$ are determined by
$\Gamma_{\alpha,0}^{(M)}(E_n^{(\alpha)})=0$.
\goodbreak

\glnonull\PLUS               
\section{Summary and discussion}
In this paper I have presented a path integral approach for the
incorporation of two- and three dimensional $\delta$-function
perturbations by means of a perturbation expansion. The motivation
was to demonstrate the validity and applicability of the formalism in
the context of path integrals, i.e.\ ``to build up quantum mechanics
from the point of view of fluctuating paths [\DK]'', hence to put the
results of the operator and the path integral approach in quantum
mechanics on an equal footing, and to supply steps towards a complete
classification of solvable Feynman path integrals [\GRSf]. A
regularization procedure was needed to treat appearing divergencies,
but could be systematically discussed. As was shown, a regularization
procedure known from functional analysis could be applied in the context
of path integrals. In order to do this, the coupling $\gamma$ in the
heuristic expression ``$\gamma\delta(\vec x-\vec a)$'' in the Lagrangian
in the path integral had to be properly interpreted and it was shown
that the formal series summation (\numba) of the perturbative path
integral approach (\numbd) corresponding to the Hamiltonian (\numbh)
could be regularized for two and three dimensional $\delta$-function
perturbations, making the two and three dimensional analogue of
(\numba) well-defined. We succeeded in developing the theory intaking
into account potential problems together with $\delta$-function
perturbations in two and three dimensions. The general feature of the
Green function of the perturbed problem with one point-interaction at
$\vec x=\vec a$ ($\vec x',\vec x'',\vec a\in\bbbr^2,\bbbr^3$) has the
form
\plus$$G^{(\delta)}(\vec x'',\vec x';E)
  =G^{(V)}(\vec x'',\vec x';E)
  +\big(\Gamma_{\alpha,\vec a}^{(V)}(E)\big)^{-1}
   G^{(V)}(\vec x'',\vec a;E)G^{(V)}(\vec a,\vec x';E)
  \tag\NUM.\num$$
with $G^{(V)}(E)$ the Green function of the unperturbed one. The
energy eigenvalues $E_n$ are determined by
\plus$$\Gamma_{\alpha,\vec a}(E_n^{(\alpha)})=0\enspace,
  \tag\NUM.\num$$
giving in general a transcendental expression for the $E_n^{(\alpha)}$.
Generally, $\alpha<0$ generates bound states, whereas $\alpha>0$
generates resonance states.

We demonstrated the formalism with a sample of two- and
three-dimensional examples. They included the free particle, the
harmonic oscillator, the Coulomb potential case, and some potential
models familiar in nuclear and electronic shell physics. The latter were
studied in three-dimensional space for $s$-waves only. We chose the
potential step, the Wood-Saxon potential, and the rotating Morse
oscillator. Of course, the Wood-Saxon potential is just but one model
in the class of potentials related to the modified P\"oschl-Teller
potentials, other exactly solvable potentials can also be considered,
e.g.\ the modified P\"oschl-Teller potential itself, or more
specifically the Hulth\'en or a Scarf-like potential. These latter
potentials can be used to describe a screened Coulomb potential, where
an additional $\delta$-function perturbation at the center would model
a point interaction (of the nucleus) in the environment of an overall
Coulomb-screening, say, of an electron shell, respectively, a point
interaction (inside a nucleus) in the environment of an overall
screening of the strong force in the nucleus itself, and many more.

One may ask, what happens in higher dimensions? Actually, point
interactions do not make sense for $D\geq4$: ``Any possible
mathematical definition of a self-adjoint operator $H$ of the heuristic
form $-\Delta+\lambda\delta_y$ in $L^2(\bbbr^d)$ should take into
account the fact that, on the space $C_0^\infty(\bbbr^d\setminus\{0\})$
of smooth functions which vanish outside a compact subset on the
complement of $\{y\}$ in $\bbbr^d$, $H$ should coincide with $-\Delta$.
For $d\geq4$ this already forces $H$ to be equal to $-\Delta$ on
$H^{2,2}(\bbbr^d)$ since $-\Delta\vert_{C_0^\infty(\bbbr^d\setminus\{0\}
)}$ is essentially self-adjoint for $d\geq4$ [\AGHHb, p.2]'', and c.f.\
[\RS, Theorem X.11, p.161]. What remains are $\delta$-function
perturbations on planes and hyperplanes, respectively, along
perpendicular lines and planes [\GROw], systems which can model a
Casimir effect, e.g.\ [\BHR].

In one dimension, making the strength of the $\delta$-function
perturbation infinitely repulsive, produces Dirichlet
boundary-conditions at the location of the $\delta$-function
perturbation. In two and three dimensions, in setting $\alpha=0$ some
sort of defect is produced, which one may call a zero-radius hard core.
Let us emphasize that again the entire Green function has to be taken
into account and not only part of it (compare [\KUSRI]). This shows
once more that it is not enough just to throw away the continuous part
of the spectrum (say, in the Coulomb case), but a proper regularization
always takes into account the entire Green function yielding
inadmissible contributions from both the discrete and continuous
spectrum of the unperturbed problem. This concludes the discussion.

\vfill
\ack{I would like to thank L.~Dabrowski (Trieste) and M.~Bordag
(Leipzig) for fruitful discussions concerning the subject of
$\delta$-function perturbations in two and three dimensions.}

\glnonull\Chapno=1           
\appendix{Green function for the harmonic oscillator}
We want to demonstrate how to calculate the Green function of the
harmonic oscillator. Of course, it is always possible to determine
the Green function of a quantum mechanical problem by means of an
operator approach exploiting the theory of differential equations.
However, this is not the point of view here. The purpose here is to
show how a Green function can be obtained by knowing the propagator
form a path integral calculation.

We consider the propagator $K(x'',x';T)$ (\numci) of the harmonic
oscillator. We need the integral representation ([\BUC, p.86; \GRA,
p.729], $a_1>a_2$, $\Re(\half+\mu-\nu)>0$):
\plus$$\multline
  \int_0^\infty\coth^{2\nu}{x\over2}\exp\bigg[
                               -{a_1+a_2\over2}t\cosh x\bigg]
  I_{2\mu}(t\sqrt{a_1a_2}\sinh x)dx
  \\
  ={\Gamma(\half+\mu-\nu)\over t\sqrt{a_1a_2}\,\Gamma(1+2\mu)}
  W_{\nu,\mu}(a_1t)M_{\nu,\mu}(a_2t)\enspace.
  \endmultline
  \tag\AA.\num$$
\edef\numAa{\AA.\num}%
Here $W_{\nu,\mu}(z)$ and $M_{\nu,\mu}(z)$ denote Whittaker-functions.
Furthermore we make use of the relations of the parabolic cylinder
functions in terms of Whittaker-functions [\BUC, pp.39]
\plus$$\myalign
  D_\nu(z)
  &=2^{\nu/2}\bigg({z^2\over2}\bigg)^{-1/4}
   W_{\nu/2+1/4,\pm1/4}\bigg({z^2\over2}\bigg)
  \tag\AA.\num a\\
  E_\nu^{(0)}(z)
  &=\sqrt{2}\,\e^{-z^2/4}{_1}
           F_1\bigg(-\nu/2;\half;{z^2\over2}\bigg)
   =\sqrt{2\pi}\,\bigg({z^2\over2}\bigg)^{-1/4}
    \CM_{\nu/2+1/4,-1/4}\bigg({z^2\over2}\bigg)
  \\   &
  \tag\AA.\num b\\
  E_\nu^{(1)}(z)
  &=\sqrt{2}\,\e^{-z^2/4}{_1}
           F_1\bigg({1-\nu\over2};{3\over2};{z^2\over2}\bigg)
   =\sqrt{2\pi}\,\bigg({z^2\over2}\bigg)^{-1/4}
  \CM_{\nu/2+1/4,+1/4}\bigg({z^2\over2}\bigg)\enspace.
  \\   &
  \tag\AA.\num c\endalign$$
\edef\numAb{\AA.\num}%
We now obtain
\plus$$\myalign
  &G(x'',x';E)
  \\   &
  =\ih\int_0^\infty dT\,\e^{\i ET/\hbar}
   \bigg({m\omega\over2\pi\i\hbar\sin\omega T}\bigg)^{1/2}
   \exp\bigg\{{\i m\omega\over2\hbar}
    \bigg[({x'}^2+{x''}^2)\cot\omega T
   -2{x'x''\over\sin\omega T}\bigg]\bigg\}
  \\   &
 \hbox{(expand the $x'x''$ term in the exponential into modified
        Bessel-functions)}
  \\   &
  ={m\omega\sqrt{x'x''}\over2\hbar^2}
   \int_0^\infty{dT\over\sin\omega T}
   \exp\bigg[-{m\omega\over2\i\hbar}({x'}^2+{x''}^2)\cot\omega T
       +{\i ET\over\hbar}\bigg]
  \\   &\qquad\times
   \left[I_{1/2}\bigg({m\omega x'x''\over\i\hbar\sin\omega T}\bigg)
   +I_{-1/2}\bigg({m\omega x'x''\over\i\hbar\sin\omega T}\bigg)\right]
  \\   &
  \hbox{(perform coordinate transformation and Wick-rotation)}
  \\   &
  ={m\sqrt{x'x''}\over2\hbar^2}\int_0^\infty dv
  \bigg(\coth{v\over2}\bigg)^{E/\hbar\omega}
  \e^{-m\omega({x'}^2+{x''}^2)\cosh v/2\hbar}
  \\   &\qquad\times
   \left[I_{1/2}\bigg({m\omega\over\hbar}x'x''\sinh v\bigg)
     +I_{-1/2}\bigg({m\omega\over\hbar}x'x''\sinh v\bigg)\right]
  \\   &
  \hbox{(apply (\numAa), set $\nu=-\half+E/\hbar\omega$)}
  \\   &
  ={1\over2\omega\hbar\sqrt{x'x''}}
     W_{\nu/2+1/4,1/4}\bigg({m\omega\over\hbar}x_>^2\bigg)
  \\   &\qquad\times
   \bigg[\Gamma\bigg({1-\nu\over2}\bigg)
   \CM_{\nu/2+1/4,1/4}\bigg({m\omega\over\hbar}x_<^2\bigg)
   +\Gamma\bigg(-\nu/2\bigg)
  \CM_{\nu/2+1/4,-1/4}\bigg({m\omega\over\hbar}x_<^2\bigg)\bigg]
  \\   &
  \hbox{(apply (\numAb))}
  \\   &
  =\sqrt{m\over\pi\hbar^3\omega}
    \Gamma\bigg({E\over\hbar\omega}-\half\bigg)
  D_{-1/2+E/\hbar\omega}\left(\sqrt{2m\omega\over\hbar}\,
   x_>\right)
  D_{-1/2+E/\hbar\omega}\left(-\sqrt{2m\omega\over\hbar}\,
   x_<\right)\enspace,
  \\   &
  \tag\AA.\num\endalign$$
which is the result of (\numcj).
The particular structure of the propagator allows to derive
the higher dimensional case from the lower one.
Let $\vec x\in\bbbr^D$ and define $\mu={\vec x'}^2+{\vec x''}^2$,
$\nu=\vec x'\cdot \vec x''$, then the following relation
for the Feynman kernel for the harmonic oscillator is valid
$$\myalign
  K^{(D)}(\vec x'',\vec x';T)
  &={1\over2\pi}{\partial\over\partial\nu}
   K^{(D-2)}(\vec x'',\vec x';T)
  \tag\AA.\num\\   \global\plus
  &={1\over2\pi}\e^{-\i\omega T}
   \bigg({m\omega\over\hbar}-2{\partial\over\partial\mu}\bigg)
   K^{(D-2)}(\vec x'',\vec x';T)\enspace.
  \tag\AA.\num\endalign$$
Introducing $\xi=\half(\vert\vec x'+\vec x''\vert
+\vert\vec x''-\vec x'\vert)$, $\eta=\half(\vert\vec x'+\vec x''\vert
-\vert\vec x''-\vec x'\vert)$ we obtain for dimensions $D=1,3,5\hdots$
for the Green function
\plus$$\myalign
  \ih \int_0^\infty  dT\,&\e^{\i TE/\hbar}
  \int\limits_{\vec x(t')=\vec x'}^{\vec x(t'')=\vec x''}\CD^D\vec x(t)
  \exp\left[{\i m\over2\hbar}
  \int_{t'}^{t''}({\dot{\vec x}}^2-\omega^2{\vec x}^2)dt\right]
         \\   &
  =\sqrt{m\over\pi\hbar^3\omega}\,
     \Gamma\bigg(\half-{E\over\hbar\omega}\bigg)
  \bigg({1\over2\pi}\bigg)^{D-1\over2}\bigg[{1\over\eta^2-\xi^2}
  \bigg(\eta{\partial\over\partial\xi}-\xi{\partial\over\partial\eta}
  \bigg)\bigg]^{D-1\over2}
         \\   &\qquad\times
  D_{-\half+{E\over\hbar\omega}}\left(\sqrt{2m\omega\over\hbar}\,
   \xi\right)
  D_{-\half+{E\over\hbar\omega}}\left(-\sqrt{2m\omega\over\hbar}\,
   \eta\right)\enspace.
  \tag\AA.\num\endalign$$
Note the various sign conventions used in the literature [\BAVE].

\baselineskip=12 truept
\glnonull\PLUS               
\appendix{Propagator in two dimensions}
We consider (\numca). The first term is just the free particle
propagator in two dimensions. To treat the second term we consider
the following inverse Laplace transformation table [\EMOTb]:

\medskip
$$\aligned
  &\vbox{\eightpoint\eightrm
         \offinterlineskip
\halign{&\vrule#&
   $\strut\ \hfil#\hfil\ $\cr
\noalign{\hrule}
height2pt&\omit&&\omit&\cr
&g(s)=\dsize{\int_0^\infty} f(t)\e^{-st}dt
    &&f(t)
    &\cr
height2pt&\omit&&\omit&\cr
\noalign{\hrule}
\noalign{\hrule}
height2pt&\omit&&\omit&\cr
&g(as)
      &&\dsize{1\over a} f\bigg(\dsize{t\over a}\bigg)
      &\cr
height2pt&\omit&&\omit&\cr
\noalign{\hrule}
height2pt&\omit&&\omit&\cr
&g_1(s)g_2(s)
      &&\dsize{\int_0^t} f_1(u)f_2(t-u)du
      &\cr
height2pt&\omit&&\omit&\cr
\noalign{\hrule}
height2pt&\omit&&\omit&\cr
&\dsize{1\over\ln s}
      &&\dsize{\int_0^\infty}\dsize{du\over\Gamma(u)}t^{u-1}
      &\cr
height2pt&\omit&&\omit&\cr
\noalign{\hrule}
height2pt&\omit&&\omit&\cr
&K_\nu(a\sqrt{s}\,)K_\nu(b\sqrt{s}\,)
      &&\dsize{1\over2t}\e^{-(a^2+b^2)/4t}
         K_\nu\bigg(\dsize{ab\over2t}\bigg)
      &\cr
height2pt&\omit&&\omit&\cr
\noalign{\hrule}
}} \endaligned$$
\medskip\noindent
Obviously, we will get a rather complicated expression, so that we keep
the discussion short. We use imaginary time $\tau=\i T$, which
has the consequence that we must set $-E\to s$. Furthermore we introduce
the abbreviations
\plus$$a={\sqrt{2m}\over\hbar}\vert\vec x''-\vec a\vert\enspace,\qquad
  b={\sqrt{2m}\over\hbar}\vert\vec x'-\vec a\vert\enspace,\qquad
  \beta=\sqrt{m\over2\hbar^2}\,\e^{\pi\alpha\hbar^2/m-\Psi(1)}
  \enspace.
  \tag\BB.\num$$
Applying successively the above rules we obtain for the propagator
of the free particle perturbed by a point interaction in $\bbbr^2$
(compare also [\ABD])
\plus$$\multline
  K^{(\delta)}(\vec x'',\vec x';\tau)
  ={m\over2\pi\hbar\tau}\,\exp\bigg[-{m\over2\hbar\tau}
       (\vec x''-\vec x')^2\bigg]
         \\
  +{m\over\pi\beta^2\hbar^2}\int_0^\infty{dv\over\Gamma(v)}
   \int_0^\tau{du\over u}
   \bigg({\tau-u\over\beta^2}\bigg)^{v-1}
   \e^{-(a^2+b^2)/4u}K_0\bigg({ab\over2u}\bigg)\enspace.
  \endmultline
  \tag\BB.\num$$

\bigskip\noindent
{\bf References}
\bigskip
\eightpoint\eightrm
\def\refno{\item}
\baselineskip=9.5 truept
\refno{[\ABD]}
S.Albeverio, Z.Brz\'ezniak and L.Dabrowski:
Fundamental Solution of the Heat and Schr\"odinger Equations with
Point Interaction; {\it Bochum University preprint}, 1993.
\refno{[\AGHHa]}
S.Albeverio, F.Gesztesy, R.J.H\o egh-Krohn and H.Holden:
Point Interactions in Two Dimensions: Basic Properties, Approximations
and Applications to Solid State Physics;
{\it J.Reine Angew.Math.}\ {\bf 380} (1987) 87
\refno{[\AGHHb]}
S.Albeverio, F.Gesztesy, R.J.H\o egh-Krohn and H.Holden:
Solvable Models in Quantum Mechanics
(Springer, Berlin-Heidelberg, 1988)
\refno{[\AGSTA]}
M.P.Avakian, G.S.Pogosyan, A.N.Sissakian and V.M.Ter-Antonyan:
Spectroscopy of a Singular Linear Oscillator;
{\it Phys.Lett.}\ {\bf 124} (1987) 233;
\newline
A.N.Sissakian, V.M.Ter-Antonyan, G.S.Pogosyan and I.V.Lutsenko:
Supersymmetry of a One-Dimensional Hydrogen-Atom;
{\it Phys.Lett.}\ {\bf 143} (1990) 247;
Three Views of the Problem of Degeneration in One-Dimensional
Quantum Mechanics;
in ``5.~International Symposium on Selected Topics in Statistical
Physics'', Dubna, 1989, p.512
({\it World Scientific}, Singapore, 1990)
\refno{[\BAVE]}
V.L.Bakhrakh and S.I.Vetchinkin:
Green's Functions of the Schr\"odinger Equation for the Simplest
Systems;
{\it Theor.Math.Phys.}\ {\bf 6} (1971) 283
\refno{[\BVK]}
V.L.Bakhrakh, S.I.Vetchinkin and S.V.Khristenko:
Green's Function of a Multidimensional Isotropic Harmonic Oscillator;
{\it Theor.Math.Phys.}\ {\bf 12} (1972) 776
\refno{[\BAU]}
D.Bauch:
The Path Integral for a Particle Moving in a $\delta$-Function
Potential;
{\it Nuovo Cimento} {\bf B 85} (1985) 118
\refno{[\BF]}
F.A.Berezin and L.D.Faddeev:
A Remark on Schr\"odinger's equation with a Singular Potential;
{\it Soviet Math.Dokl.}\ {\bf 2} (1961) 372
\refno{[\BLWE]}
J.M.Blatt and V.F.Weisskopf:
Theoretical Nuclear Physics
(Springer, New York-Heidel\-berg-Berlin, 1979)
\refno{[\BOVO]}
M.Bordag and S.Voropaev:
Charged Particle with Magnetic Moment in the Aharonov-Bohm Potential;
{\it Leipzig University preprint 1993} NTZ-93-07
\refno{[\BHR]}
M.Bordag, D.Henning and D.Robaschik:
Quantum Field Theory With External Potentials Concentrated on Planes;
{\it J.Phys.A: Math.Gen.}\ {\bf 25} (1992) 4483
\refno{[\BUGE]}
W.Bulla and F.Gesztesy:
Deficiency Indices and Singular Boundary Conditions in Quantum
Mechanics;
{\it J.Math.Phys.}\ {\bf 26} (1985) 2520
\refno{[\BUC]}
H.Buchholz: The Confluent Hypergeometric Function,
{\it Springer Tracts in Natural Philosophy, Vol.15}
(Springer, Berlin-Heidelberg, 1969)
\refno{[\CIWI]}
P.Y.Cai, A.Inomata and R.Wilson:
Path-Integral Treatment of the Morse Oscillator;
{\it Phys.Lett.}\ {\bf A 96} (1983) 117
\refno{[\CAR]}
M.Carreau:
The Functional Integral for a Free Particle on a Half-Plane;
{\it J.Math.Phys.}\ {\bf 33} (1992) 4139
\refno{[\CFG]}
M.Carreau, E.Farhi and S.Gutmann:
Functional Integral for a Free Particle in a Box;
{\it Phys.Rev.}\ {\bf D 42} (1990) 1194
\refno{[\CHc]}
L.Chetouani and T.F.Hammann:
Coulomb Green's Function, in an $n$-Dimensional Euclidean Space;
{\it J.Math.Phys.}\ {\bf 27} (1986) 2944
\refno{[\CMS]}
T.E.Clark, R.Menikoff and D.H.Sharp:
Quantum Mechanics on the Half-Line Using Path Integrals;
{\it Phys.Rev.}\ {\bf D 22} (1980) 3012
\refno{[\DAGR]}
L.Dabrowski and H.Grosse:
On Nonlocal Point Interactions in One, Two and Three Dimensions;
{\it J.Math.Phys.}\ {\bf 26} (1985) 2777
\refno{[\DURa]}
I.H.Duru:
Morse-Potential Green's Function With Path Integrals;
{\it Phys.Rev.}\ {\bf D 28} (1983) 2689
\refno{[\DK]}
I.H.Duru and H.Kleinert:
Solution of the Path Integral for the H-Atom;
{\it Phys.Lett.}\ {\bf B 84} (1979) 185;
Quantum Mechanics of H-Atoms From Path Integrals;
{\it Fort\-schr.Phys.}\ {\bf 30} (1982) 401
\refno{[\EMOTb]}
A.Erd\'elyi, W.Magnus, F.Oberhettinger and F.G.Tricomi (eds.):
Tables of Integral Transforms, Vol.I (McGraw Hill, New York, 1954)
\refno{[\FH]}
R.P.Feynman and A.Hibbs: Quantum Mechanics and Path Integrals,
(McGraw Hill, New York, 1965)
\refno{[\FLU]}
S.Fl\"ugge:
Practical Quantum Mechanics, Vol.I;
Die Grundlagen der mathematischen Wis\-sen\-schaf\-ten Vol.177
(Springer, Berlin-Heidelberg, 1977)
\refno{[\GASCH]}
B.Gaveau and L.S.Schulman:
Explicit Time-Dependent Schr\"odinger Propagators;
{\it J.Phys.A: Math.Gen.}\ {\bf 19} (1986) 1833
\refno{[\GOOb]}
M.J.Goovaerts:
Path-Integral Evaluation of a Nonstationary Calogero Model;
{\it J.Math.Phys.}\ {\bf 16} (1975) 720
\refno{[\GODE]}
M.J.Goovaerts, A.Babceno, and J.T.Devreese:
A New Expansion Method in the Feynman Path Integral Formalism:
Application to a One-Dimensional Delta-Function Potential;
{\it J.Math.Phys.}\ {\bf 14} (1973) 554;
\newline
M.J.Goovaerts and J.T.Devreese:
Analytic Treatment of the Coulomb Potential in the Path Integral
Formalism by Exact Summation of a Perturbation Expansion;
{\it J.Math.Phys.}\ {\bf 13} (1972) 1070;
and Erratum: {\it J.Math.Phys.}\ {\bf 14} (1973) 153
\refno{[\GOBR]}
M.J.Goovaerts and F.Broeckx:
Analytic Treatment of a Periodic $\delta$-Function Potential in the
Path Integral Formalism;
{\it SIAM J.Appl.Math.}\ {\bf 45} (1985) 479
\refno{[\GRA]}
I.S.Gradshteyn and I.M.Ryzhik:
Table of Integrals, Series, and Products,
(Academic Press, New York, 1980)
\refno{[\GROb]}
C.Grosche:
The Path Integral on the Poincar\'e Upper Half-Plane With
a Magnetic Field and for the Morse Potential;
{\it Ann.Phys.(N.Y.)} {\bf 187} (1988) 110
\refno{[\GROe]}
C.Grosche:
Path Integral Solution of a Class of Potentials Related to the
P\"oschl-Teller Potential;
{\it J.Phys.A: Math.Gen.}\ {\bf 22} (1989) 5073
\refno{[\GROh]}
C.Grosche:
Path Integrals for Potential Problems With $\delta$-Function
Perturbation;
{\it J.Phys.A: Math.Gen.}\ {\bf 23} (1990) 5205
\refno{[\GROm]}
C.Grosche:
Coulomb Potentials by Path-Integration;
{\it Fortschr.Phys.}\ {\bf 40} (1992) 695
\refno{[\GROs]}
C.Grosche:
Path Integral Solution of Scarf-Like Potentials;
{\it Trieste preprint}, SISSA\-/179\-/92\-/FM, October 1992
\refno{[\GROu]}
C.Grosche:
Selberg Trace-Formul\ae\ in Mathematical Physics;
Conference Proceedings Vol.41 of the Workshop ``From Classical to
Quantum Chaos (1892-1992)'', Trieste, 21-24 July, 1992, p.45,
eds.: G.\ Dell'Antonio, S.\ Fantoni and V.\ R.\ Manfredi
({\it Societ\`a Italiana Di Fisica}, Bologna, 1993)
\refno{[\GROw]}
C.Grosche:
$\delta$-Function Perturbations and Boundary Problems by Path
Integration;
{\it Ann.Physik} {\bf 2} (1993)
\refno{[\GROx]}
C.Grosche:
Path Integration via Summation of Perturbation Expansions and
Application to Totally Reflecting Boundaries and Potential Steps;
{\it Phys.Rev.Lett.}\ {\bf 71} (1993) 1
\refno{[\GRSb]}
C.Grosche and F.Steiner:
Path Integrals on Curved Manifolds;
{\it Zeitschr.Phys.}\ {\bf C 36} (1987) 699
\refno{[\GRSf]}
C.Grosche and F.Steiner,
Classification of Solvable Feynman Path Integrals;
{\it DESY preprint} DESY 92-189, to appear in the {\it Proceedings of
the ``Fourth International Conference on Path Integrals from $meV$ to
$MeV$'', May 1992, Tutzing, Germany} (World Scientific, Singapore)
\refno{[\GRSg]}
C.Grosche and F.Steiner:
Table of Feynman Path Integrals;
to appear in: {\it Springer Tracts in Modern Physics}
\refno{[\HEJ]}
D.A.Hejhal: The Selberg Trace Formula for $\PSL(2,\bbbr)$, I\&II;
{\it Lecture Notes in Mathematics} {\bf 548, 1001}
(Springer, Berlin-Heidelberg, 1976)
\refno{[\HOI]}
R.Ho and A.Inomata:
Exact Path Integral Treatment of the Hydrogen Atom;
{\it Phys.Rev.Lett.}\ {\bf 48} (1982) 231
\refno{[\HOST]}
L.Hostler:
Runge--Lenz Vector and the Coulomb Green's Function;
{\it J.Math. Phys.}\ {\bf 8} (1967) 642
\refno{[\INOb]}
A.Inomata:
Alternative Exact-Path-Integral-Treatment of the Hydrogen Atom;
{\it Phys.Lett.}\ {\bf A 101} (1984) 253
\refno{[\JAMA]}
R.K.Janev and Z.Mari\'c:
Perturbation of the Spectrum of Three-Dimensional Harmonic Oscillator
by a $\delta$-Potential;
{\it Phys.Lett.}\ {\bf A 46} (1974) 313
\refno{[\BJ]}
G.Junker and M.B\"ohm:
The $\SU(1,1)$ Propagator as a Path Integral Over Noncompact Groups;
{\it Phys.Lett.}\ {\bf A 117} (1986) 375;
Path Integration Over Compact and Noncompact Rotation Groups;
{\it J.Math.Phys.}\ {\bf 28} (1987) 1978
\refno{[\KLEh]}
H.Kleinert:
How to do the Time Sliced Path Integral for the H Atom;
{\it Phys.Lett.}\ {\bf A 120} (1987) 361
\refno{[\KLEMUS]}
H.Kleinert and I.Mustapic:
Summing the Spectral Representations of P\"oschl-Teller and
Rosen-Morse Fixed-Energy Amplitudes;
{\it J.Math.Phys.}\ {\bf 33} (1992) 643
\refno{[\KOM]}
I.V.Komarov:
Application of the Short Range Potential in the Calculations of the
Ion-Ion Recombination; {\it Sixth International Conference on the
Physics of Electronics and Atomic Collisions}, Massachusetts, 1969,
abstract of papers, p.1015
\refno{[\KUSRI]}
N.Kumar and R.Sridhar:
Exact Energy Eigenvalues for an Attractive Coulomb Potential with
Zero-Radius Hard Core;
{\it Phys.Lett.}\ {\bf A 39} (1972) 389
\refno{[\LABH]}
S.V.Lawande and K.V.Bhagwat:
Feynman Propagator for the $\delta$-Function Potential;
{\it Phys.Lett.}\ {\bf A 131} (1988) 8
\refno{[\MANO]}
E.B.Manoukian:
Explicit Derivation of the Propagator for a Dirac Delta Potential;
{\it J.Phys.A: Math.Gen.}\ {\bf 22} (1989) 67
\refno{[\MERZ]}
E.Merzbacher:
Quantum Mechanics
(J.Wiley \&\ Sons, New York, $1970^2$)
\refno{[\PAKSa]}
K.Pak and I.S\"okmen:
A New Exact Path Integral Treatment of the Coulomb and the Morse
Potential Problems;
{\it Phys.Lett.}\ {\bf A 100} (1984) 327
\refno{[\PI]}
D.Peak and A.Inomata:
Summation Over Feynman Histories in Polar Coordinates;
{\it J.Math.Phys.}\ {\bf 10} (1969) 1422
\refno{[\PT]}
G.P\"oschl und E.Teller:
Bemerkungen zur Quantenmechanik des anharmonischen Oszillators;
{\it Zeitschr.Phys.}\ {\bf 83} (1933) 143
\refno{[\PRAN]}
R.E.Prange:
Quantized Hall Resistance and the Measurement of the Fine-Structure
Constant;
{\it Phys.Rev.}\ {\bf B 23} (1981) 4802
\refno{[\RS]}
M.Reed and B.Simon:
Methods of Modern Mathematical Physics. II. Fourier Analysis, Self
Adjointness,
(Academic Press, New York, 1975)
\refno{[\RUND]}
J.Rundgren:
The Rotating Morse Oscillator;
{\it Arkiv Fysik} {\bf 30} (1965) 61
\refno{[\SCATE]}
S.Scarlatti and A.Teta:
Derivation of the Time-Dependent Propagator for the Three-Dimensional
Schr\"o\-dinger Equation with One Point Interaction;
{\it J.Phys.A: Math.Gen.}\ {\bf 23} (1990) L1033
\refno{[\SCHUe]}
L.S.Schulman:
Techniques and Applications of Path Integration,
(John Wiley \&\ Sons, New York, 1981)
\refno{[\SEL]}
A.Selberg:
Harmonic Analysis and Discontinuous Groups in Weakly Symmetric
Riemannian Spaces with Application to Dirichlet Series;
{\it J.Indian Math.Soc.}\ {\bf 20} (1956) 47
\refno{[\STEc]}
F.Steiner:
Exact Path Integral Treatment of the Hydrogen Atom;
{\it Phys.Lett.}\ {\bf A 106} (1984) 363
\refno{[\VEN]}
A.B.Venkov:
Spectral Theory of Automorphic Functions;
{\it Proc.Math.Inst.Steklov} {\bf 153} (1981) 1


\enddocument